%% file: sdp.tex
\documentclass{article}
\usepackage{arxiv}

\usepackage[utf8]{inputenc} 
\usepackage[T1]{fontenc}    
\usepackage{hyperref}       
\usepackage{url}            
\usepackage[nodayofweek]{datetime} 
\usepackage{booktabs}       
\usepackage{tabularx}       
\usepackage{amsfonts}       
\usepackage{amssymb}        
\usepackage{nicefrac}       
\usepackage{microtype}      
\usepackage{graphicx}
\usepackage{natbib}
\usepackage{doi}
\usepackage{xcolor}

\usepackage{amsmath}
\usepackage{amsthm}
\newtheorem{theorem}{Theorem}[section]
\newtheorem{corollary}{Corollary}[theorem]
\newtheorem{lemma}[theorem]{Lemma}
\newtheorem{proposition}{Proposition}[section]
\newtheorem{definition}{Definition}[section]
\newtheorem{example}{Example}[section]

\newtheorem{axiom}{Axiom}[section]

\newtheorem{observation}{Observation}[section]

\renewcommand{\include}[1]{\input{#1}}

\newtheorem{sdpassumption}[axiom]{Assumption}

\newtheorem{sdpalgorithm}[observation]{Algorithm}

\counterwithout{corollary}{theorem}\counterwithin{corollary}{section}
\makeatletter
\AtBeginDocument{\let\sdp@qed\qed
  \renewcommand{\qed}{\ifvmode\vspace{-\parskip}\noindent\hfill\qedsymbol\par\else\sdp@qed\fi}}
\makeatother

\definecolor{sdplink}{RGB}{128,0,64}
\definecolor{sdpcite}{RGB}{0,70,140}
\hypersetup{colorlinks = true,
linkcolor = sdplink,
urlcolor  = sdpcite,
citecolor = sdpcite,
anchorcolor = black}
\usdate

\title{Theoretical Foundations of Sequential Dynamic Programming}

\newdate{articleDate}{20}{10}{2026}
\date{\displaydate{articleDate}}

\makeatletter
\let\@fnsymbol\@arabic
\makeatother

\author{\href{https://orcid.org/0000-0002-5937-8905}{\includegraphics[scale=0.06]{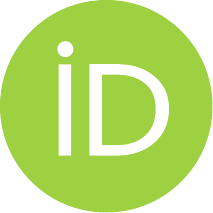}}\hspace{1mm}Alan Lujan\footnotemark[1]\\
Johns Hopkins University\\}

\renewcommand{\headeright}{}
\renewcommand{\undertitle}{}
\renewcommand{\shorttitle}{Sequential Dynamic Programming}

\hypersetup{
pdftitle={Theoretical Foundations of Sequential Dynamic Programming},
pdfsubject={},
pdfauthor={Alan Lujan},
pdfkeywords={dynamic programming,sequential decomposition,intermediate states,sufficient statistics,separability,curse of dimensionality},
addtopdfcreator={Written in Curvenote}
}

\begin{document}
\maketitle
\footnotetext[1]{Correspondence to: alujan@jhu.edu}

\begin{abstract}
Many economic problems have several controls per period, whose joint choice is subject to the curse of dimensionality. Solving them in stages is computationally cheaper, but no general result says when staging is exact. I develop a separability theory, discrete choices and non-concave objectives included: staging is exact if utility, transitions, and constraints depend on earlier decisions only through a state passed forward. For each ordering the smallest such state exists and is unique up to relabeling. Any state that keeps staging exact for all preferences over later decisions and the future must carry everything the smallest one carries for the transition. For affine primitives, the smallest state consists of the linear combinations, such as cash-on-hand, on which later outcomes depend. Its dimensions pick the order whose largest stage is smallest. In a six-choice household model, each stage chooses one variable on a three-dimensional state, the joint problem all six.
\end{abstract}

\keywords{dynamic programming, sequential decomposition, intermediate states, sufficient statistics, separability, curse of dimensionality}

\include{intro-intro}

\include{intro-theory}

\include{intro-characterization}

\include{intro-discovering}

\include{intro-two-asset}

\include{intro-conclusion}

\include{intro-appendix}

\include{intro-supplement}

\section*{Acknowledgments}
I thank Christopher D. Carroll, Matthew N. White, Akshay Shanker, and Sebastian Benthall for helpful discussions over many years. During the preparation of this work I used Claude Code (Anthropic) to assist in editing and revising the prose and in checking the proofs and the numerical claims. After using this tool, I reviewed and edited the content as needed and take full responsibility for the content of the publication.

\bibliographystyle{plainnat-sdp}
\bibliography{main.bib}

\end{document}

%% file: intro-intro.tex
\section{Introduction}\label{sec:introduction}

A household can choose its hours and its consumption at once, or hours first and then consumption given the resources its labor has produced. Where first-order conditions cannot replace search (as with a discrete choice or a non-concave objective), the second procedure is computationally cheaper by orders of magnitude. With a risky share added, a search over three controls on grids with 100 points per dimension already costs $100^3 = 10^6$ evaluations per state before any other source of dimensionality is counted, whereas a sequence of stage problems searches only over each stage's block of controls. Yet practice generally solves the joint form: in the incomplete-markets economies descended from \citet{riskfree_huggett_1993}, \citet{uninsured_aiyagari_1994}, and \citet{income_krusell_1998}, and just as much in structural life-cycle models of consumption, labor supply, and human capital \citep{intertemporal_imai_2004}, a period with several decisions is handled by simultaneous first-order conditions, intratemporal substitutions, or direct search over the whole control vector. Staging appears only at the margins, where a method guarantees its validity by construction: the nested endogenous grid method solves the final consumption choice inside the search over the others \citep{guide_druedahl_2021}, problems whose first-order conditions form a triangular system are solved one condition at a time \citep{multidimensional_iskhakov_2015}, and the sequential endogenous grid method (Sequential EGM) inverts stage by stage \citep{sequential_lujan_2026}. Carrying the whole history always makes the two forms agree, yet far less can suffice, as cash-on-hand has shown since \citet{saving_deaton_1991} and \citet{bufferstock_carroll_1997} by collapsing assets and income into a single state. But no general result says what the carried variables must be.

Where staging is attempted, the answers come from intuition, case by case. Intuition usually works, but where it fails, a staged solver still returns policies, and nothing in its output shows that they are wrong. Each staging method in use guarantees validity only for the class of problems it was built for; none characterizes from the primitives what another problem's stages must carry, or whether that state is smaller than the history it replaces. The human-capital model of \citet{intertemporal_imai_2004} shows what goes wrong without such a characterization. \citet{multidimensional_iskhakov_2015} classified it, by the structure of its first-order conditions, as an application of their triangular method; their correction \citep{multidimensional_corrigendum_2017} later reversed that classification. I show that the model decomposes nonetheless, so the structure of the first-order conditions is an unreliable guide to decomposability. I classify problems by their primitives instead, whatever form the first-order conditions take. I call a Markov decision process (MDP) whose uncertainty resolves between periods \textit{sequentially decomposable} when its transition, utility, and constraints all decompose through a common chain of intermediate states, each carrying what the period's earlier decisions leave for the later ones. The within-period optimization then splits into stage problems over parts of the control vector. I show that sequential decomposability gives the two forms the same solution, then characterize from the primitives the smallest state each stage must carry.

Call two early histories (each the state together with the decisions already taken in the period) equivalent when they produce identical outcomes under \textit{every} possible \textit{completion}, a choice of the period's remaining decisions. Identifying equivalent histories yields the minimal intermediate state, which always exists: it is the history's \textit{outcome profile}, the function sending each completion to its outcome. In the motivating example, a history is the wealth brought into the period, the wage, and the hours just chosen. Its outcome profile maps consumption and the risky share to the post-decision state, the saving and risky share the period leaves for the next, so two histories share it exactly when they leave the same cash-on-hand. Resting only on the budget identity, the argument holds as written when hours take finitely many values and so covers a participation margin, which first-order conditions cannot handle. No valid state carries less, because every valid intermediate state determines the outcome profile. The minimal state after one decision, together with the next decision, determines the minimal state after it. For a fixed ordering of decisions the chain of minimal states is unique up to relabeling. But across orderings it is not unique. Finding the minimal state once for each set of earlier decisions gives the number of variables each stage handles under every ordering, and so picks the order whose largest stage is smallest.

Swapping a dollar of the household's wealth for a dollar of labor income leaves cash-on-hand, and so every later outcome, unchanged. At a fixed wage the budget is affine in the history, the form budget constraints usually take. For affine transitions the minimal state is a set of linear aggregates, one for each independent direction of change in the history that changes some later outcome. The swap is the one direction that changes no later outcome, so the state needs only one dimension, cash-on-hand, whatever the wage. When the constraints and the utility terms that interact with later decisions are affine as well, an explicit algorithm builds from the primitives a state that suffices for the transition, the utility and the constraints at once. Conversely, a state that keeps staging exact for every preference over later decisions, and for every valuation of the future in a family rich enough to tell any two outcomes apart, must carry everything the transition needs. That part of the minimal state is therefore determined by the transition, whatever the preferences. With the model's utility added to the test preferences and remaining choices kept feasible, the same holds for utility and constraints.

Some primitives outside the affine class still admit a state smaller than the history. Suppose the marginal effects of later decisions depend on the history only through a statistic that the modeler can name and that those effects pin down. Then the statistic, together with the primitives evaluated at one reference choice of the later decisions for each value of the discrete ones, gives the minimal state. In the six-choice household of Section~\ref{sec:two-asset-example}, the minimal state after hours collapses liquid wealth, participation and hours into cash-on-hand and carries the illiquid account and the house unchanged. The states found after successive decisions assemble into a sequential decomposition, with each stage's utility and feasible set built from the primitives, provided the states vary continuously with the history, including its exogenous states, and no feasible choice disappears abruptly.

Classical separability theory studies the two-block case of what a stage must carry: when a function of two variable groups factors through an aggregate of the first, from \citet{separability_sono_1961} and \citet{separability_leontief_1947} through \citet{topological_debreu_1960} and \citet{structure_gorman_1968} to the treatise of \citet{duality_blackorby_1978}, with the two-stage budgeting of \citet{utilitytree_strotz_1957} and \citet{separable_gorman_1959} as its canonical economic application. I inherit the question and depart in two ways. First, the dimension of the aggregate is an output. Those authors ask whether a group of variables fixed in advance admits an aggregator, one scalar per group, whereas the minimal intermediate state can have any dimension. I compute it by linear algebra when the primitives are affine, or from a statistic their marginal effects pin down. \citet{latent_blundell_2000} come closest, selecting the number of latent separable groups in demand systems. Second, the separability is \textit{joint}: one chain of minimal states must serve transition, utility, and constraints at once. Stacking three separate aggregates, where each exists, produces such a chain, though at a cost in dimension. A swap of histories must now be invisible to the transition, the utility and the constraints at once; the joint state can exceed any primitive's own aggregate.

The minimal state is to a period's decisions what a minimal sufficient statistic is to a likelihood. With utility and constraints included, two histories are identified when every completion gives them the same outcome and feasible set and their utilities differ by a term constant across completions, much as two samples are when their log-likelihoods differ by a term free of the parameter \citep{completeness_lehmann_1950}. The minimal sufficient statistic of a history, with its existence and uniqueness, is classical outside economics: the Nerode relation of automata theory \citep{linear_nerode_1958}, the minimal model of a Markov decision process \citep{equivalence_givan_2003}, and the minimal realization of a control system (the smallest state space reproducing an input-output map), whose dimension is a rank for linear systems (\citet{mathematical_kalman_1963}; \citet{effective_ho_1966}) and, under constant rank, for nonlinear ones (\citet{existence_sussmann_1976}; \citet{nonlinear_hermann_1977}). In economics, \citet{finite_kalai_1988} measure a repeated-game strategy's complexity by the number of distinct strategies it induces after histories, which equals the number of states of its smallest automaton. This paper concerns what that statistic becomes in an economic model, where a single state serves all three primitives. Staged optimization has an older literature: \citet{composition_mitten_1964} identified monotone aggregation as sufficient for recursive solution, \citet{standardform_witsenhausen_1973} gave sequential stochastic control a staged normal form, and the post-decision state at the center of \citet{approximate_powell_2011} is the one used here. \citet{dynamic_sargent_2025} refactor the Bellman operator into expectation, discounting and maximization and iterate on expected values or Q-factors of state-action pairs; here the action itself splits into stages, each with a state constructed from the primitives. \citet{space_sanjurjo_2026} counts the working memory that an order of reading attributes requires and shows that finishing one alternative before the next minimizes it. His count depends on how computation is represented; the minimal state here is fixed by the primitives up to relabeling.

The staged solvers this theory justifies descend from the endogenous grid method of \citet{endogenous_carroll_2006}, which turns a first-order condition into an assignment by inverting it on a grid of post-decision states. Its descendants push the same inversion through discrete choices, non-convexities, multiple assets, higher-dimensional grids, and stages without closed-form inverses \citep{generalization_barillas_2007, generalized_fella_2014, endogenous_iskhakov_2017, general_druedahl_2017, endogenous_ludwig_2018, endogenous_hallengreen_2024}. Sequential EGM takes as given exactly the intermediate states this paper constructs. \citet{endogenous_white_2015} states conditions for EGM over a period as a whole: a post-decision state through which the transition factors and from which feasibility can be read, with the controls invertible in it. Sequential decomposability requires transition, utility and feasibility to factor at every intermediate state, with no invertibility condition. The nearest method, the Triangular Dynamic Optimization of \citet{multidimensional_iskhakov_2015}, requires the first-order-condition system to be globally triangular and analytically invertible. Sequential decomposability restricts only the primitives.

Section~\ref{sec:theoretical_foundations} sets up the problem, shows when staging reproduces the joint solution, and shows what every valid state must carry. Section~\ref{sec:algebraic_characterization} characterizes the minimal state, Section~\ref{sec:discovering} shows how to discover it, Section~\ref{sec:two-asset-example} works a household example with six choices, and Section~\ref{sec:conclusion} names the open problem. Proofs of the main results are in the Appendix. The Supplemental Appendix holds the remaining proofs, the examples' details, and the Imai-Keane decomposition.

%% file: intro-theory.tex
\section{A Theory of Within-Period Decomposability}\label{sec:theoretical_foundations}

Before formalizing, consider a household in the manner of \citet{labor_bodie_1992} that enters the period with wealth $b$ and a wage $\theta$, and then chooses hours $n$, consumption $c$, and the share $\varsigma \in [0, 1]$ of its saving $a = b + \theta n - c$ held in a risky asset. The wage and the return are i.i.d. and independent of each other. Next period's wealth is $(R_f + \varsigma(R' - R_f))a$, where the risky gross return $R'$ arrives after every decision of the period and $R_f$ is the safe one. Here the control vector is $\mathbf{A} = (n, c, \varsigma)$. It is natural to choose hours first, consumption second, and the portfolio last. Hours produce cash-on-hand $m = b + \theta n$. The utility and constraints of the later stages depend on wealth, the wage and hours only through $m$, so the wage leaves the state after the first stage. Consumption then depends on $m$ alone and leaves the saving $a$. The post-decision state is $(a, \varsigma)$: the period must carry the share until the return, realized between periods, combines the risky total $\varsigma a$ with the safe total $(1 -\varsigma)a$. At zero saving every share leaves the same law, so the shares are identified there; the map to $(a, \varsigma a)$ embeds the identified space in the plane. Neither $m$ nor $a$ is a state or control; the theory derives both from the primitives.

\subsection{The High-Dimensional Dynamic Programming Problem}\label{sec:hdp_problem}

The high-dimensional dynamic programming (HDP) problem is an MDP whose action is a vector of controls. Throughout, $\mathcal{B}(\mathcal{Z})$ denotes the Banach space of bounded continuous functions on a complete metric space $\mathcal{Z}$, with the supremum norm; all functions, sets, and policies satisfy standard measurability conditions \citep{recursive_stokey_1989}.

\begin{definition}[The HDP Problem]\label{def-hdp}Maximize the expected discounted sum of utilities, starting from an initial state $\mathbf{X}_0$:
\begin{equation}
V(\mathbf{X}_0) = \sup_{\{\mathbf{A}_t\}_{t=0}^\infty} \mathbb{E}_0 \left[\sum_{t=0}^\infty \beta^t \,U(\mathbf{X}_t,\mathbf{A}_t)\right],
\end{equation}
where the discount factor $\beta \in (0,1)$ governs time preference and the supremum runs over policies adapted to the information available when each choice is made, subject to time-invariant dynamics:

\begin{itemize}
\item $\mathbf{Y}_t = F(\mathbf{X}_t,\mathbf{A}_t)$ \textit{(post-decision state)}
\item $\mathbf{X}_{t+1} \sim G(\mathbf{Y}_t)$ \textit{(law of motion)}
\item $\mathbf{A}_t \in \Gamma(\mathbf{X}_t)$ \textit{(feasibility correspondence)}
\end{itemize}

\end{definition}The state $\mathbf{X}_t$ includes every state variable, endogenous or exogenous, persistent or i.i.d.: wealth and a house alongside productivity or a transitory wage, all realized between periods and seen before the period's decisions. In the examples, $\varepsilon$ denotes a persistent (Markov) exogenous state and $\theta$ a transitory (i.i.d.) one. The law of motion $G$ is the standard stochastic transition of a Markov decision process: at the end of the period it draws next period's state from the post-decision state. Through $G$, a risky return realized after the period's decisions enters next period's wealth without becoming a state of its own, and it may covary with next period's exogenous states. I take the post-decision state $\mathbf{Y}_t$ to hold what the period's choices and states leave for the future, and only that: two pairs $(\mathbf{X}_t, \mathbf{A}_t)$ with $\mathbf{A}_t \in \Gamma(\mathbf{X}_t)$ give the same $\mathbf{Y}_t$ exactly when they give the same law of next period's state. The post-decision state carries a persistent exogenous state whenever its distinct values predict different next values. An i.i.d. state such as a transitory wage predicts nothing. When next period's state depends on it only through what the choices produced, as with the wage through wealth, $F$ does not list it, and the stages drop it once no later primitive depends on it. The within-period theory takes the continuation value as given, so it applies period by period to finite-horizon problems with age-dependent primitives, where backward induction replaces the contraction; I state the infinite-horizon case (Theorem~\ref{prop-hdp-solution}).

\begin{axiom}[HDP Regularity]\label{ass-hdp}The primitives satisfy:

\begin{enumerate}
\item The state space $\mathcal{X}$ is nonempty and compact, a convex subset of $\mathbb{R}^{n_x}$ times a finite set of values of discrete coordinates. The control space is a product $\mathcal{A} = \mathcal{A}^{(1)} \times \cdots \times \mathcal{A}^{(n_a)}$ of scalar control ranges, each a compact interval for a continuous control or a finite set for a discrete one.\footnote{The product structure is imposed on $\mathcal{A}$, not on $\Gamma(\mathbf{X})$, so feasibility still couples the controls. It makes the domains $\mathcal{E}_j$ and $\mathcal{L}_j$ of Section~\ref{sec:algebraic_characterization} compact and defines $F$ and $U$ at every recombination $(\mathbf{e}_j, \mathbf{l}_j)$ that the minimal-state construction and the Necessity Theorem evaluate.}
\item The period utility $U: \mathcal{X} \times \mathcal{A} \to \mathbb{R}$ is continuous and bounded. The transition $F: \mathcal{X} \times \mathcal{A} \to \mathbb{R}^{n_Y}$ is continuous, with $\mathcal{Y}$ its image, and the law of motion $G$ is a transition function from $\mathcal{Y}$ to $\mathcal{X}$ with the Feller property.
\item The feasibility correspondence $\Gamma$ is continuous with nonempty compact values.\footnote{The Supplemental Appendix shows one-sided utility bounds suffice, projects next period's state into $\mathcal{X}$ off feasible plans, and weakens continuity of $\Gamma$ to closed graphs for eligibility thresholds.}
\end{enumerate}

\end{axiom}These are the standard conditions of \citet{recursive_stokey_1989} and \citet{dynamic_bertsekas_2017}, from which Theorem~\ref{prop-hdp-solution} derives a unique solution. Convexity in the continuous coordinates lets the Supplemental Appendix repair, by projection, a formula for next period's state that, at infeasible plans, takes values outside $\mathcal{X}$. As a product, $\mathcal{A}$ splits into compact blocks under every ordered partition of the controls. I call a control vector $\mathbf{A} \in \Gamma(\mathbf{X})$ a \textit{feasible plan} for the period. Below, ``the constraints'' means $\Gamma$ and its stage counterparts, and \textit{constraint functions} $\zeta_k$ are any representation $\Gamma(\mathbf{X}) = \{\mathbf{A} : \zeta_k(\mathbf{X}, \mathbf{A}) \geq 0 \text{ for all } k\}$, equalities written as pairs of inequalities. Where an equality can be solved for one variable, as a budget identity can for end-of-period assets, I solve it into the transition.

No smoothness or concavity is assumed. Only the results of Section~\ref{sec:discovering} and their extensions in the Supplemental Appendix use differentiability, in continuous variables, so kinked budgets such as progressive tax schedules remain within the Equivalence Theorem and the results of Section~\ref{sec:algebraic_characterization}. Strict concavity adds uniqueness to the Equivalence Theorem, which holds for sets of optimal plans.\footnote{Primitive sufficient conditions for strict concavity are standard (\citet{recursive_stokey_1989}, Theorem 4.8).} Smoothness of $V$, and so of the continuation value $W$, does not follow from the conditions above, nor does any result in the main text use it.

\subsection{The Bellman Equation}

The \textit{Bellman operator} $\mathcal{T}$ acts on $\mathcal{B}(\mathcal{X})$ by
\begin{equation}
(\mathcal{T}v)(\mathbf{X}) = \max_{\mathbf{A} \in \Gamma(\mathbf{X})} \left\{ U(\mathbf{X}, \mathbf{A}) + \beta \int v \, dG\bigl(F(\mathbf{X}, \mathbf{A})\bigr) \right\};
\end{equation}
a function $V \in \mathcal{B}(\mathcal{X})$ satisfies the \textit{Bellman equation} if $V = \mathcal{T}V$.

\begin{theorem}[Existence and Uniqueness of the HDP Value Function]\label{prop-hdp-solution}Under Assumption~\ref{ass-hdp}, the Feller property of $G$ makes $\mathbf{Y} \mapsto \int v \, dG(\mathbf{Y})$ continuous for bounded continuous $v$. By the Theorem of the Maximum the Bellman operator then maps $\mathcal{B}(\mathcal{X})$ into itself, where it satisfies Blackwell's sufficient conditions and is a contraction with modulus $\beta$, so a unique $V$ with $V = \mathcal{T}V$ exists (\citet{recursive_stokey_1989}, Theorems 3.6 and 3.3). Bounded returns and $\beta < 1$ make that fixed point the value of the sequence problem of Definition~\ref{def-hdp}, and the maximizers in $\mathcal{T}V$, a nonempty, compact-valued, upper hemicontinuous correspondence, have measurable selections, each an optimal stationary policy (\citet{stochastic_bertsekas_1978}, Propositions 9.10 and 9.12; \citet{recursive_stokey_1989}, Theorem 7.6).

\end{theorem}The \textit{continuation value function} $W(\mathbf{Y}) = \beta \int V \, dG(\mathbf{Y})$ is the expected value of starting next period from the post-decision state; like $V$, it is unique, bounded, and continuous. With it, the Bellman equation reads
\begin{equation}
V(\mathbf{X}) = \max_{\mathbf{A}\in\Gamma(\mathbf{X})} \bigl\{ U(\mathbf{X}, \mathbf{A})+W(F(\mathbf{X}, \mathbf{A}))\bigr\}.
\end{equation}
Everything that follows concerns the within-period problem on the right-hand side, a single maximization over the full control vector with $W$ fixed. Throughout, I assume that all uncertainty resolves between periods, with the period's exogenous states realized before its decisions and next period's state drawn after the last of them. Many models draw a shock between two decisions, as when a household sets its hours before a wage shock and chooses consumption after it. But splitting the period at each shock gives a sequence of subperiods, each carrying its post-decision state across the next shock. Each subperiod's decisions then pose a within-period problem of the kind studied here, with the next subperiod's expected value in place of the continuation value. Under expected utility the assumption is thus without loss of generality: with $K$ subperiods, a discount factor $\beta^{1/K}$ and subperiod $k$'s utility scaled by $\beta^{ -k/K}$ reproduce the model's discounting, and the subperiod index joins the state.

\subsection{The Sequential Dynamic Programming Structure}\label{sec:sp_structure}

Sequential Dynamic Programming (SDP) replaces the joint maximization with stage problems whenever the period has the structure defined here.

Consider an ordered partition of controls $\mathbf{A} = (\mathbf{a}_1, \ldots, \mathbf{a}_J)$ where $\mathbf{a}_j \in \mathcal{A}_j \subseteq \mathbb{R}^{n_{a_j}}$. \textit{Cut} $j$ is the passage from stage $j -1$ to stage $j$. It divides the period's variables into the early block $(\mathbf{X}, \mathbf{a}_1, \ldots, \mathbf{a}_{j -1})$ and the late controls $(\mathbf{a}_j, \ldots, \mathbf{a}_J)$. Stages may hold several controls each; whether a partition decomposes depends on the primitives. A decomposition has four components, defined in turn, the first describing how each stage's controls transform the state within the period.

\begin{definition}[Stage Transition Function]\label{def-stage-transition}A \textit{stage transition function} $f_j$ maps the current intermediate state $\mathbf{y}_{j -1}$ and the stage controls $\mathbf{a}_j$ to the next intermediate state $\mathbf{y}_{j} = f_j(\mathbf{y}_{j -1}, \mathbf{a}_j) \in \mathcal{Y}_j$. It is defined on the \textit{generated} pairs $(\mathbf{y}_{j -1}, \mathbf{a}_j)$, those with $\mathbf{a}_j \in \mathcal{A}_j$ and $\mathbf{y}_{j -1}$ the intermediate state of some history, feasible or not; that state uses only $f_1, \ldots, f_{j -1}$.

\end{definition}\begin{definition}[Intermediate State]\label{def-intermediate-state}The \textit{intermediate state} $\mathbf{y}_j$ after stage $j$ is a function of the history $(\mathbf{X}, \mathbf{a}_1, \ldots, \mathbf{a}_j)$. It lies in a compact metric space $\mathcal{Y}_j$,\footnote{Its dimension $n_j$, where finite, is the covering dimension, which is $n_j$ for a subset of $\mathbb{R}^{n_j}$ with nonempty interior (\citet{dimension_hurewicz_1941}, Chapter IV and Theorem V 8); the Supplemental Appendix explains why the spaces of minimal states qualify as stage spaces.} and
\begin{equation}
\mathbf{y}_0 = \mathbf{X}, \quad \mathbf{y}_{j} = f_j(\mathbf{y}_{j-1}, \mathbf{a}_j) \text{ for } j = 1, \ldots, J.
\end{equation}

\end{definition}\begin{definition}[Stage Utility Function]\label{def-stage-utility}A \textit{stage utility function} $u_j$, real-valued on the generated pairs of stage $j$, captures the direct utility contribution from stage $j$, depending only on the entering intermediate state and the stage's controls. A stage without direct utility (for example, pure portfolio reallocation) has $u_j \equiv 0$.

\end{definition}\begin{definition}[Stage Feasibility Correspondence]\label{def-stage-constraint}A \textit{stage feasibility correspondence} $\gamma_j: \mathcal{P}_{j -1} \rightrightarrows \mathcal{A}_j$ collects the controls satisfying the constraints of stage $j$ given the entering intermediate state. Its domain is the set $\mathcal{P}_{j -1} \subseteq \mathcal{Y}_{j -1}$ of \textit{reachable} states, the values at histories $(\mathbf{X}, \mathbf{a}_1, \ldots, \mathbf{a}_{j -1})$ that some later controls complete to a plan in $\Gamma(\mathbf{X})$. It is compact, as a continuous image of a compact set.

\end{definition}\begin{definition}[Intermediate Value Function]\label{def-intermediate-value}The \textit{intermediate value function} $v_j: \mathcal{P}_{j -1} \to \mathbb{R}$ represents the maximum attainable value from stage $j$ onward, given the entering intermediate state $\mathbf{y}_{j -1}$:
\begin{equation}
v_j(\mathbf{y}_{j-1}) = \max_{\mathbf{a}_j \in \gamma_j(\mathbf{y}_{j-1})} \left\{ u_j(\mathbf{y}_{j-1}, \mathbf{a}_j) + v_{j+1}(f_j(\mathbf{y}_{j-1}, \mathbf{a}_j)) \right\}
\end{equation}
with terminal condition $v_{J+1}(\mathbf{y}_{J}) \equiv W(\mathbf{y}_{J})$. Under (S3) below, every control in $\gamma_j(\mathbf{y}_{j -1})$ leads to a reachable state, where $v_{j+1}$ is defined. Continuity of $u_j$ and $f_j$, with $\gamma_j$ nonempty, continuous, and compact-valued, then makes each maximum attained and, by Berge's maximum theorem applied backward from the continuous $v_{J+1} = W$, each $v_j$ continuous, so the recursion is well defined.

\end{definition}\begin{definition}[Sequential Decomposability]\label{def-sequential-decomposability}A \textit{sequential decomposition} of an HDP with primitives $(U, F, \Gamma)$ along the ordered partition $\mathbf{A} = (\mathbf{a}_1, \dots, \mathbf{a}_J)$ is a collection $\{\mathcal{Y}_j, f_j, u_j, \gamma_j\}_{j=1}^J$ of intermediate state spaces, with $\mathcal{Y}_0 = \mathcal{X}$ and $\mathcal{Y}_{J} = \mathcal{Y}$, continuous stage transitions, continuous bounded stage utilities, and stage feasibility correspondences continuous and compact-valued on the reachable states, such that, for all $(\mathbf{X}, \mathbf{A}) \in \mathcal{X} \times \mathcal{A}$, feasible or not, the intermediate states $\mathbf{y}_j = \mathbf{y}_j(\mathbf{X}, \mathbf{a}_1, \ldots, \mathbf{a}_j)$ satisfy

(S1) \textit{Transition consistency.} $F(\mathbf{X}, \mathbf{A}) = \mathbf{y}_{J}$;

(S2) \textit{Utility stage additivity.} $U(\mathbf{X}, \mathbf{A}) = \sum_{j=1}^J u_j(\mathbf{y}_{j -1}, \mathbf{a}_j)$;

(S3) \textit{Constraint sequential compatibility.} For each $j$, the history $(\mathbf{X}, \mathbf{a}_1, \ldots, \mathbf{a}_j)$ completes to a plan in $\Gamma(\mathbf{X})$ if and only if $(\mathbf{X}, \mathbf{a}_1, \ldots, \mathbf{a}_{j -1})$ does and, at its then reachable state, $\mathbf{a}_j \in \gamma_j(\mathbf{y}_{j -1})$. Chained over $j = 1, \ldots, J$, (S3) says that $\mathbf{A} \in \Gamma(\mathbf{X})$ exactly when every stage choice lies in its stage set along the plan's states.

The HDP is \textit{sequentially decomposable} along the partition if it admits a sequential decomposition along it, and \textit{decomposable on feasible plans} if some such collection satisfies (S3) as stated and (S1) and (S2) at every feasible plan, $\mathbf{A} \in \Gamma(\mathbf{X})$.

\end{definition}Condition (S3) makes every stage set nonempty at the reachable states, its only domain, since a history that completes to a feasible plan has a completion whose stage-$j$ choice lies in $\gamma_j$. The recursion evaluates stage sets only there.

In the motivating example the three stages start from $\mathbf{y}_0 = (b, \theta)$. The first has the transition $f_1(b, \theta, n) = b + \theta n = m$ and labor disutility $u_1(b, \theta, n) = -\varphi n^{1+\nu}/(1+\nu)$, the second $f_2(m, c) = m - c = a$ and consumption utility $u_2(m, c) = c^{1 -\sigma}/(1 -\sigma)$, and the third $f_3(a, \varsigma) = (a, \varsigma)$, $u_3 = 0$ and $\gamma_3(a) = [0, 1]$. Counting state and control coordinates, the joint problem has five continuous dimensions and no stage more than three. I keep $u_2$ bounded with a consumption floor $\underline{c} > 0$; the Supplemental Appendix gives $\gamma_1$, $\gamma_2$ and the restrictions that keep every state feasible.

Sequential decomposability asks for the \textit{joint} separability of all three primitives through a \textit{common} chain of intermediate states. Taken alone at cut $j$, (S1) extends the Leontief-Sono condition of weak separability from a scalar aggregate to one of any dimension, whose scalar differential form is the Goldman-Uzawa test \citep{note_goldman_1964}. Condition (S2) nests weak separability and additive separability across the two groups. The closest antecedents of the joint requirement that I know of (the conditions of \citet{conditions_gorman_1968} and \citet{additively_debreu_1982}) describe a single function.

Every HDP satisfies (S1)-(S3) under the one-stage partition $J = 1$, with nothing to verify. Call a decomposition \textit{nontrivial} if $J \geq 2$, and \textit{compressive} if the largest stage dimension $\overline{d} = \max_j \{n_{j -1} + n_{a_j}\}$ falls short of the joint dimension $n_0 + n_a$, where $n_0 = n_x$ counts the continuous coordinates of the state. In every dimension count, a stage's state is counted on its reachable states, the stage problem's domain, and $n_a$ and $n_{a_j}$ count continuous controls only, since a discrete control multiplies a stage's computational cost by its number of values. Here ``decomposes'' always means ``decomposes nontrivially,'' along a stated ordered partition or over the finite set of candidate orders. The Supplemental Appendix states when compression lowers the worst-case computational cost exponent from $n_0 + n_a$ to at most $\overline{d}$.

\subsection{The Equivalence Theorem}

\begin{theorem}[Equivalence of HDP and SDP under Sequential Structure]\label{prop-sdp-hdp-equivalence}Let $V$ be the unique value function of the HDP problem (Definition~\ref{def-hdp}, Theorem~\ref{prop-hdp-solution}) satisfying Assumption~\ref{ass-hdp}. If this HDP problem is also sequentially decomposable, or only decomposable on feasible plans (Definition~\ref{def-sequential-decomposability}), then the value function $v_1$ resulting from the recursive maximization (Definition~\ref{def-intermediate-value}), evaluated at the initial state $\mathbf{y}_0 = \mathbf{X}$, is identical to $V$:
\begin{equation}
V(\mathbf{X}) = v_1(\mathbf{X}) \quad \text{for all } \mathbf{X} \in \mathcal{X}.
\end{equation}
A plan $\mathbf{A}^* = (\mathbf{a}_{1}^*, \ldots, \mathbf{a}_{J}^*)$ maximizes the HDP's within-period objective $U + W \circ F$ over $\Gamma(\mathbf{X})$ if and only if, along its own intermediate states $\mathbf{y}_j^* = f_j(\mathbf{y}_{j -1}^*, \mathbf{a}_j^*)$ from $\mathbf{y}_0^* = \mathbf{X}$, each stage choice solves its stage problem,
\begin{equation}
\mathbf{a}_{j}^* \in \arg\max_{\mathbf{a}_j \in \gamma_j(\mathbf{y}_{j-1}^*)} \bigl\{u_j(\mathbf{y}_{j-1}^*, \mathbf{a}_j) + v_{j+1}\bigl(f_j(\mathbf{y}_{j-1}^*, \mathbf{a}_j)\bigr)\bigr\}.
\end{equation}
If $\Gamma(\mathbf{X})$ is convex and $U + W \circ F$ strictly concave in $\mathbf{A}$ on it, that plan is unique.

\end{theorem}The proof, given in the Appendix, applies the principle of optimality \citep{dynamic_bellman_1957} by backward induction over the stages, recursively \textit{within} the decision period rather than across periods. Because the proof evaluates only feasible plans, decomposition on feasible plans suffices. It can admit a smaller state than discovery finds (Supplemental Appendix).

The results extend to Epstein-Zin preferences \citep{substitution_epstein_1989, dynamic_ren_2020} through a transformed reward that can enlarge the minimal state, though risk aversion never changes it (Supplemental Appendix).

\subsection{Necessity: Serving All Test Preferences Forces Factorization}\label{sec:necessity}

A state that supports staging must be a sufficient statistic for the transition when it serves a family of test preferences. To test a candidate state, we pin the late controls to an arbitrary target with a steep quadratic penalty. As the penalty steepens, the stage value converges to the continuation value of the transition at that target. Any family of continuation values rich enough to tell post-decision states apart then recovers the transition itself.

\begin{definition}[Point-Separating Family]\label{def-point-separating}A family $\mathcal{W} \subseteq \mathcal{B}(\mathcal{Y})$ of continuation value functions \textit{separates points} of $\mathcal{Y}$ if for every pair $\mathbf{y} \neq \mathbf{y}'$ in $\mathcal{Y}$ there exists $W \in \mathcal{W}$ with $W(\mathbf{y}) \neq W(\mathbf{y}')$.

\end{definition}Point separation is a weak requirement: with $\underline{y}^{(i)}$ the least value of the $i$th coordinate $y^{(i)}$ on the compact $\mathcal{Y} \subseteq \mathbb{R}^{n_Y}$, the shifted log-linear family $\{\sum_i \lambda_i \log(1 + y^{(i)} - \underline{y}^{(i)}) : \boldsymbol{\lambda} \in \mathbb{R}^{n_Y}_{++}\}$ consists of bounded, strictly concave, strictly increasing functions and separates points, whatever the sign of the post-decision state. Taken over all bounded continuous $V$, the continuation values $W$ they induce separate the post-decision states of feasible plans, whose laws of next period's state differ. Off the feasible plans a law of motion repaired by projection can merge post-decision states, so there a family such as the log-linear one is needed to separate them. Linear functions of next period's state depend only on its mean and need not separate even feasible ones.

\begin{theorem}[Necessity of Transition Factorization]\label{thm-necessity}Let the HDP satisfy Assumption~\ref{ass-hdp}, let $(\mathbf{a}_1, \mathbf{a}_2)$ be a two-stage partition of the controls, let $g$ be a continuous map from $\mathcal{X} \times \mathcal{A}_1$ to $\mathcal{Y}_1$, and let $\mathcal{W}$ be any point-separating family (Definition~\ref{def-point-separating}). The following are equivalent:

(i) $F$ factors through $g$: some function $\tilde{F}$, defined and continuous on the image of $g$ times $\mathcal{A}_2$, satisfies $F(\mathbf{X}, \mathbf{a}_1, \mathbf{a}_2) = \tilde{F}(g(\mathbf{X}, \mathbf{a}_1), \mathbf{a}_2)$ for all $(\mathbf{X}, \mathbf{a}_1, \mathbf{a}_2)$.

(ii) For every quadratic penalty weight $M > 0$, target $\mathbf{a}_2^{\circ} \in \mathcal{A}_2$ and $W \in \mathcal{W}$, the conditional value
\begin{equation}
\Lambda(\mathbf{X}, \mathbf{a}_1) := \max_{\mathbf{a}_2 \in \mathcal{A}_2} \left\{ -M\|\mathbf{a}_2 - \mathbf{a}_2^{\circ}\|^2 + W(F(\mathbf{X}, \mathbf{a}_1, \mathbf{a}_2)) \right\}
\end{equation}
depends on $(\mathbf{X}, \mathbf{a}_1)$ only through $g(\mathbf{X}, \mathbf{a}_1)$.

(iii) For every bounded continuous stage utility $u_2: \mathcal{Y}_1 \times \mathcal{A}_2 \to \mathbb{R}$ and every bounded continuous $W: \mathcal{Y} \to \mathbb{R}$, the conditional value defined as in (ii) with $u_2(g(\mathbf{X}, \mathbf{a}_1), \mathbf{a}_2)$ in place of the penalty depends on $(\mathbf{X}, \mathbf{a}_1)$ only through $g(\mathbf{X}, \mathbf{a}_1)$.

The result extends to $J$-stage partitions by applying the two-stage statement at each cut, with the early variables $(\mathbf{X}, \mathbf{a}_1, \ldots, \mathbf{a}_{j -1})$ as the first block and the late controls $(\mathbf{a}_j, \ldots, \mathbf{a}_J)$ as the second, whose compactness Assumption~\ref{ass-hdp} already supplies. This delivers factorization cut by cut, whereas assembling the cut-wise statistics into a chain, and so into (S1), is the separate work of Lemma~\ref{lem-chaining}.

\end{theorem}The proof is in the Appendix. One continuation value cannot replace the family: a continuation value constant where $F$ differs separates nothing there.

Cash-on-hand thus summarizes early choices for every objective sharing the budget transition, given (S2) for the objective and (S3) for the constraints.

%% file: intro-characterization.tex
\section{Sufficient Statistics and the Characterization of Decomposability}\label{sec:algebraic_characterization}

Throughout this section and Section~\ref{sec:discovering}, fix a stage $j$ and write $\mathbf{e}_j$ for the ``early variables'' $(\mathbf{X}, \mathbf{a}_1, \ldots, \mathbf{a}_{j -1})$, with domain $\mathcal{E}_j = \mathcal{X} \times \mathcal{A}_1 \times \cdots \times \mathcal{A}_{j -1}$, and $\mathbf{l}_j = (\mathbf{a}_j, \ldots, \mathbf{a}_J)$ for the ``late controls,'' with domain $\mathcal{L}_j = \mathcal{A}_j \times \cdots \times \mathcal{A}_J$. Both domains are compact by Assumption~\ref{ass-hdp}. Exogenous states are early variables like any other, and the minimal state keeps only the part of them on which the later primitives depend. In the results that use interiors of $\mathcal{E}_j$, or derivatives or affinity in the early variables, the continuous exogenous coordinates count among those variables. The discrete early variables enter as parameters, and so does any exogenous coordinate in which the primitives lack the required affinity or smoothness. In the notation of Definition~\ref{def-sequential-decomposability}, the statistic that summarizes $\mathbf{e}_j$ is the incoming intermediate state $\mathbf{y}_{j -1}$; as in Definition~\ref{def-intermediate-state}, $n_{j -1}$ denotes its dimension where finite.

\begin{definition}[Sufficient Statistic Equivalence]\label{def-sufficient-statistic-equivalence}For a transition function $F: \mathcal{E}_j \times \mathcal{L}_j \to \mathcal{Y}$, define the \textit{sufficient statistic equivalence relation} $\sim_j$ on $\mathcal{E}_j$ by:
\begin{equation}
\mathbf{e}_j \sim_j \mathbf{e}'_j \quad \iff \quad F(\mathbf{e}_j, \mathbf{l}_j) = F(\mathbf{e}'_j, \mathbf{l}_j) \text{ for all } \mathbf{l}_j \in \mathcal{L}_j,
\end{equation}
so histories with different exogenous states can be equivalent: the relation compares profiles only.

\end{definition}\begin{theorem}[Existence, Minimality, and Canonical Realization of the Intermediate State]\label{thm-quotient-minimality}Let $F$ be a continuous map from $\mathcal{E}_j \times \mathcal{L}_j$ to $\mathcal{Y}$ on compact domains. Define the \textit{outcome profile map}
\begin{equation}
\Phi_j: \mathcal{E}_j \to C(\mathcal{L}_j, \mathcal{Y}), \qquad \Phi_j(\mathbf{e}_j) = F(\mathbf{e}_j, \cdot),
\end{equation}
where $C(\mathcal{L}_j, \mathcal{Y})$, the continuous maps $\mathcal{L}_j \to \mathcal{Y}$, has the supremum metric. Then:

\begin{enumerate}
\item \textit{Existence.} $\Phi_j$ is continuous, $\mathbf{e}_j \sim_j \mathbf{e}'_j$ if and only if $\Phi_j(\mathbf{e}_j) = \Phi_j(\mathbf{e}'_j)$, and $F$ factors through $\Phi_j$: there is a continuous $\tilde{F}$ with $F(\mathbf{e}_j, \mathbf{l}_j) = \tilde{F}(\Phi_j(\mathbf{e}_j), \mathbf{l}_j)$. The image $\mathcal{Q}_j := \Phi_j(\mathcal{E}_j)$ is a compact metric space.
\item \textit{Terminality.} If $g: \mathcal{E}_j \to \mathcal{Z}$ is any sufficient statistic, meaning $g(\mathbf{e}_j) = g(\mathbf{e}'_j)$ implies $F(\mathbf{e}_j, \cdot) = F(\mathbf{e}'_j, \cdot)$, then there exists a unique map $\tilde{\Phi}: g(\mathcal{E}_j) \to \mathcal{Q}_j$ with $\Phi_j = \tilde{\Phi} \circ g$. Every sufficient statistic determines the outcome profile.
\item \textit{Uniqueness.} Any two sufficient statistics with the terminality property of part 2 are related by a bijection, and by a homeomorphism if both are continuous maps on $\mathcal{E}_j$ with values in Hausdorff spaces.
\end{enumerate}

\end{theorem}Sequential decomposability requires the stages to \textit{compose}, in that the intermediate state at stage $j+1$ must be computable from the intermediate state at stage $j$ and the stage-$j$ control. Arbitrary cut-wise sufficient statistics need not compose; the minimal ones do, which lets each cut be checked alone.

\begin{lemma}[Chaining of Minimal States]\label{lem-chaining}Fix an ordered partition $(\mathbf{a}_1, \ldots, \mathbf{a}_J)$, and write $\mathcal{Q}_{J+1}$ for the image of $F$, the profile space at the empty completion. Then the minimal sufficient statistics compose: for each $j$ there is a continuous map $\tilde{f}_j: \mathcal{Q}_j \times \mathcal{A}_j \to \mathcal{Q}_{j+1}$ with
\begin{equation}
\Phi_{j+1}(\mathbf{e}_j, \mathbf{a}_j) = \tilde{f}_j\bigl(\Phi_j(\mathbf{e}_j), \mathbf{a}_j\bigr).
\end{equation}
Setting $\mathcal{Y}_0 = \mathcal{X}$ and folding $\Phi_1$ into the first transition, $f_1 = \tilde{f}_1 \circ (\Phi_1 \times \mathrm{id})$, the spaces $\mathcal{Q}_j$ of minimal states and the maps $\{\tilde{f}_j\}$ satisfy (S1), with $\mathcal{Y}_{j -1} = \mathcal{Q}_j$ for $j \geq 2$.

\end{lemma}The minimal-state construction extends from the transition to all three clauses of sequential decomposability on the histories that admit a feasible completion. Write $\mathcal{C}_j(\mathbf{e}_j)$ for the set of late controls that complete $\mathbf{e}_j$ to a feasible plan and $\mathcal{E}_j^{\mathcal{C}}$ for the set of histories with $\mathcal{C}_j(\mathbf{e}_j)$ nonempty, compact as the projection of the compact graph of $\Gamma$. Define the \textit{joint sufficiency relation} $\approx_j$ on $\mathcal{E}_j^{\mathcal{C}}$ by requiring equality of feasible completions, $\mathcal{C}_j(\mathbf{e}_j) = \mathcal{C}_j(\mathbf{e}'_j)$, and, at all late controls $\mathbf{l}_j \in \mathcal{L}_j$, feasible or not, equality of transitions $F(\mathbf{e}_j, \mathbf{l}_j) = F(\mathbf{e}'_j, \mathbf{l}_j)$ and \textit{constancy in $\mathbf{l}_j$} of the utility gap $U(\mathbf{e}_j, \mathbf{l}_j) - U(\mathbf{e}'_j, \mathbf{l}_j)$. The utility clause is weaker than equality because utility already accumulated by stage $j$ is sunk, so two histories that differ only in sunk utility support identical continuation behavior. Decomposition on feasible plans constrains just the weaker relation $\mathbf{e}_j \approx_j^{\mathcal{C}} \mathbf{e}'_j$, which imposes the transition and utility clauses only on common feasible completions $\mathcal{C}_j(\mathbf{e}_j) = \mathcal{C}_j(\mathbf{e}'_j)$.

\begin{corollary}[Minimal Joint State]\label{cor-joint-minimal}Fix a reference late control $\mathbf{l}^0 \in \mathcal{L}_j$ and write $U^0(\mathbf{e}_j) := U(\mathbf{e}_j, \mathbf{l}^0)$ for the reference utility and $U - U^0$ for the utility offset. On $\mathcal{E}_j^{\mathcal{C}}$:

\begin{enumerate}
\item $\mathbf{e}_j \approx_j \mathbf{e}'_j$ if and only if the joint profile $\Pi_j(\mathbf{e}_j)$, the triple of $\Phi_j(\mathbf{e}_j)$, $U(\mathbf{e}_j, \cdot) - U^0(\mathbf{e}_j)$ and $\mathcal{C}_j(\mathbf{e}_j)$, takes the same value at both, and parts 2 and 3 of Theorem~\ref{thm-quotient-minimality} hold with $\Pi_j$ in place of $\Phi_j$, a statistic $g$ counting as sufficient when $g(\mathbf{e}_j) = g(\mathbf{e}'_j)$ implies $\mathbf{e}_j \approx_j \mathbf{e}'_j$.
\item Viewed as a function of $\mathbf{e}_j$, the intermediate state $\mathbf{y}_{j -1}$ of any sequential decomposition is such a sufficient statistic and so determines the $\approx_j$-classes; that of a decomposition on feasible plans determines the $\approx_j^{\mathcal{C}}$-classes.
\item Conversely, let $\mathcal{W}$ be a point-separating family (Definition~\ref{def-point-separating}). A map $g$ on $\mathcal{E}_j^{\mathcal{C}}$ identifies only $\approx_j^{\mathcal{C}}$-equivalent histories if and only if there is a function $\kappa$ on $\mathcal{E}_j^{\mathcal{C}}$, the same for every test preference (each choice of $M$, $\mathbf{l}^{\circ}$ and $W$ below), such that

\begin{equation}
\max_{\mathbf{l}_j \in \mathcal{C}_j(\mathbf{e}_j)} \bigl\{ U(\mathbf{e}_j, \mathbf{l}_j) - M\|\mathbf{l}_j - \mathbf{l}^{\circ}\|^2 + W(F(\mathbf{e}_j, \mathbf{l}_j)) \bigr\} - \kappa(\mathbf{e}_j)
\end{equation}

depends on $\mathbf{e}_j$ only through $g(\mathbf{e}_j)$ for every $M > 0$, every $\mathbf{l}^{\circ} \in \mathcal{L}_j$ and every $W$ in $\mathcal{W} \cup \{0\}$.
\end{enumerate}

\end{corollary}\begin{proof}The utility clause of $\approx_j$ says that $U(\mathbf{e}_j, \cdot) - U^0(\mathbf{e}_j) = U(\mathbf{e}'_j, \cdot) - U^0(\mathbf{e}'_j)$, which gives the equivalence in statement 1. Parts 2 and 3 of the theorem use only that the profile identifies equivalent histories. For statement 2, let two histories in $\mathcal{E}_j^{\mathcal{C}}$ share $\mathbf{y}_{j -1}$. By (S1) both transitions are the same function of $\mathbf{y}_{j -1}$ and the late controls. By (S2) each utility is a term in its own early variables plus the same function of $\mathbf{y}_{j -1}$ and the late controls, so the utility gap is constant in $\mathbf{l}_j$. Each history has a feasible completion, so by (S3) its early controls satisfy $\gamma_1, \ldots, \gamma_{j -1}$, and a completion is feasible exactly when the late controls satisfy $\gamma_j, \ldots, \gamma_J$ along the states generated from $\mathbf{y}_{j -1}$; the feasible completions therefore agree. For a decomposition on feasible plans, the same steps run with $\mathbf{l}_j$ in the common $\mathcal{C}_j$. Statement 3 is proved in the Appendix.

\end{proof}Each test preference in statement 3 adds a stage-separable penalty to the model's own primitives, so sequential decomposability admits it; statement 3 is the feasible-plan counterpart of the Necessity Theorem, extended to utility and constraints on the feasible completions.

\begin{corollary}[Canonical Chain for a Fixed Order]\label{cor-canonical-chain}Fix an ordered partition of the controls. For every chain of intermediate states $\{\mathbf{y}_j\}$ satisfying (S1) at every control vector there are unique maps $\tilde{\Phi}_j$ with $\Phi_j = \tilde{\Phi}_j \circ \mathbf{y}_{j -1}$ for $j = 2, \ldots, J$, so the chain of minimal sufficient statistics is \textit{terminal}. Any two terminal chains that satisfy (S1) with continuous states are related by stage-wise homeomorphisms.

\end{corollary}\textit{Across} orders, however, uniqueness fails: the Supplemental Appendix exhibits two orders whose minimal chains do not refine each other. Since the minimal state at a cut depends only on the set of controls placed before it, discovery on these sets compares the orders from the primitives (Proposition~\ref{prop-order-search}).

%% file: intro-discovering.tex
\section{Discovering Sufficient Statistics}\label{sec:discovering}

This section computes the minimal state of Section~\ref{sec:algebraic_characterization} from the primitives, by linear algebra for affine primitives and from marginal effects beyond them, and assembles the discovered states into a decomposition.

\subsection{Affine Transitions and Known Features}\label{sec:affine_discovery}

A direction of variation in the early variables that leaves $F$ unchanged \textit{for every late control} is a direction the intermediate state may safely ignore. These directions form the \textit{common kernel}. The state must record the directions that some late control can detect. For transitions affine in the early variables the rows of the coefficient matrices span them, so linear algebra on those rows gives the minimal state. Classical separability tests require the kernel of the transition's Jacobian to be \textit{the same} for every late control. For $F(\mathbf{e}, l) = e^{(1)} + l\, e^{(2)}$ the kernel rotates with the late control and that test fails. However, the span of the coefficient rows over late controls, here all of $\mathbb{R}^2$, still gives the dimension of the state that does suffice, two. Economically, $e^{(2)}$ could be installed capital and $l$ a later choice of how intensively to use it, so resources are $e^{(1)}$ plus utilized capital. Section~\ref{sec:marginal_inversion} covers transitions whose marginals in the late controls identify a statistic of the history. For other smooth transitions the Supplemental Appendix gives a local criterion.

\begin{corollary}[Linear Minimal Sufficient Statistics for Affine Transitions]\label{cor-gradient-test}Let $F$ be affine in the early variables, $F(\mathbf{e}_j, \mathbf{l}_j) = \mathsf{A}(\mathbf{l}_j)\, \mathbf{e}_j + \mathsf{B}(\mathbf{l}_j)$, with $\mathsf{A}(\mathbf{l}_j)$ continuous in $\mathbf{l}_j$ and $\mathcal{E}_j$ of nonempty interior, which also makes $\mathsf{A}(\cdot)$ unique. Define the \textit{joint row space}
\begin{equation}
\mathcal{R} := \operatorname{span} \bigcup_{\mathbf{l}_j \in \mathcal{L}_j} \operatorname{rowspace} \mathsf{A}(\mathbf{l}_j) \subseteq \mathbb{R}^{\dim \mathbf{e}_j}, \qquad n_{j-1} := \dim \mathcal{R}.
\end{equation}
Let $\mathsf{C}$ be any $n_{j -1} \times \dim(\mathbf{e}_j)$ matrix whose rows span $\mathcal{R}$. Then:

\begin{enumerate}
\item The transition $F$ factors through the linear statistic $g(\mathbf{e}_j) = \mathsf{C}\mathbf{e}_j$: there are continuous matrices $\mathsf{D}(\mathbf{l}_j)$ with $\mathsf{A}(\mathbf{l}_j) = \mathsf{D}(\mathbf{l}_j) \mathsf{C}$, so $F = \mathsf{D}(\mathbf{l}_j)(\mathsf{C} \mathbf{e}_j) + \mathsf{B}(\mathbf{l}_j)$.
\item The $\sim_j$-classes are exactly the intersections of $\mathcal{E}_j$ with the affine subspaces $\mathbf{e}_j + \mathcal{R}^{\perp}$, and no continuous statistic of dimension smaller than $n_{j -1}$ is sufficient on $\mathcal{E}_j$ or on any subset of it with nonempty interior.
\end{enumerate}

\end{corollary}The same linear algebra applies when the transition is linear in known functions of the early variables, as it is in a Cobb-Douglas technology with capital fixed before hours are chosen and under the adjustment costs of the household model of Section~\ref{sec:two-asset-example}. Such functions exist exactly when the transition has finite rank in the late controls (its values at all late controls span a finite-dimensional space of functions of the early variables), as it does whenever every late control is discrete. The transition's values at finitely many late controls form a minimal sufficient statistic (Supplemental Appendix).

\begin{corollary}[Transitions Linear in Known Features]\label{cor-feature-linear}Let $F(\mathbf{e}_j, \mathbf{l}_j) = \mathsf{A}(\mathbf{l}_j)\, \boldsymbol{\psi}(\mathbf{e}_j) + \mathsf{B}(\mathbf{l}_j)$ for a continuous map $\boldsymbol{\psi}: \mathcal{E}_j \to \mathbb{R}^{d_\psi}$ and coefficients continuous in $\mathbf{l}_j$, and let the rows of $\mathsf{C}$ span the joint row space $\mathcal{R} \subseteq \mathbb{R}^{d_\psi}$ of $\{\mathsf{A}(\mathbf{l}_j)\}$. Then the $\sim_j$-classes are exactly the level sets of $\mathsf{C} \boldsymbol{\psi}(\mathbf{e}_j)$, so $\mathsf{C}\boldsymbol{\psi}$ is a minimal sufficient statistic in the sense of Theorem~\ref{thm-quotient-minimality}, and $F = \mathsf{A}(\mathbf{l}_j) \mathsf{C}^{+} \mathsf{C} \boldsymbol{\psi}(\mathbf{e}_j) + \mathsf{B}(\mathbf{l}_j)$ with $\mathsf{C}^{+}$ the pseudoinverse.

\end{corollary}\begin{proof}$F(\mathbf{e}_j, \cdot) = F(\mathbf{e}'_j, \cdot)$ exactly when $\mathsf{A}(\mathbf{l}_j)(\boldsymbol{\psi}(\mathbf{e}_j) - \boldsymbol{\psi}(\mathbf{e}'_j)) = 0$ for every $\mathbf{l}_j$, that is, when the difference lies in $\mathcal{R}^{\perp} = \ker \mathsf{C}$. Every row of $\mathsf{A}(\mathbf{l}_j)$ lies in $\mathcal{R}$, on which $\mathsf{C}^{+}\mathsf{C}$ is the identity.

\end{proof}The coordinates of $\mathsf{C}\boldsymbol{\psi}$ can, however, outnumber the dimension of the minimal state when the features are functionally dependent: $F(e, l) = e + l\,e^2$ has the two features $e$ and $e^2$ but a one-dimensional minimal state. Any statistic with the same level sets as $\mathsf{C}\boldsymbol{\psi}$ is then a minimal state. Applied to the stacked primitive $H$ of Section~\ref{sec:discovery}, the corollary gives the joint state under the exact-factoring convention. Because the construction depends on how the constraints are written, I write them in budget form (Supplemental Appendix).

\subsection{Primitives Identified by Their Marginals}\label{sec:marginal_inversion}

Take utility $u(c - \varphi n^{1+\nu}/(1+\nu))$ of consumption net of the disutility of hours, with $\varphi > 0$ and $u$ strictly concave, at the cut after hours. Under CRRA curvature this utility has infinite rank in consumption, yet a smaller state still suffices, because its marginal in consumption reveals hours. The budget's marginal in consumption is -1 whatever the history. Histories with equal late-control marginals have profiles that differ by a constant that one evaluation of each primitive fixes.

Let $\mathcal{D}_j$ be the finite set of values of the discrete late controls, a single point when there are none. A degenerate interval counts as a discrete control with one value. For $\mathbf{d} \in \mathcal{D}_j$ let the slice $\mathcal{L}_j^{\mathbf{d}}$ be the box of continuous late controls that accompany $\mathbf{d}$. Let $H$ be the stacked primitive of Section~\ref{sec:discovery}, or the transition $F$ alone, and write $\partial_{\mathbf{l}} H$ for its derivative in the continuous late controls. Where $H(\mathbf{e}_j, \mathbf{d}, \cdot)$ is locally Lipschitz on the interior of $\mathcal{L}_j^{\mathbf{d}}$, that derivative exists at almost every point there by Rademacher's theorem (\citet{measure_evans_2015}, Section 3.1).

\begin{proposition}[Minimal States from Late-Control Marginals]\label{prop-marginal-inversion}Let $H: \mathcal{E}_j \times \mathcal{L}_j \to \mathbb{R}^m$ be continuous, with $H(\mathbf{e}_j, \mathbf{d}, \cdot)$ locally Lipschitz on the interior of $\mathcal{L}_j^{\mathbf{d}}$ for every history $\mathbf{e}_j$ and every $\mathbf{d} \in \mathcal{D}_j$. Suppose a continuous statistic $\mathbf{s}_j: \mathcal{E}_j \to \mathbb{R}^q$ carries all the dependence of the marginal on the history: $\partial_{\mathbf{l}} H(\mathbf{e}_j, \mathbf{d}, \mathbf{l}) = \Theta_j(\mathbf{s}_j(\mathbf{e}_j), \mathbf{d}, \mathbf{l})$ for every $\mathbf{e}_j$ and $\mathbf{d}$ and almost every $\mathbf{l}$ in the interior of $\mathcal{L}_j^{\mathbf{d}}$. Fix a reference control $\mathbf{l}^{\mathbf{d}} \in \mathcal{L}_j^{\mathbf{d}}$ in each slice and set
\begin{equation}
g_j(\mathbf{e}_j) := \bigl(\mathbf{s}_j(\mathbf{e}_j),\ (H(\mathbf{e}_j, \mathbf{d}, \mathbf{l}^{\mathbf{d}}))_{\mathbf{d} \in \mathcal{D}_j}\bigr).
\end{equation}

\begin{enumerate}
\item Equal values of $g_j$ imply equal profiles $H(\mathbf{e}_j, \cdot)$: $g_j$ is sufficient.
\item Suppose also that the marginal identifies the statistic: $\Theta_j(\mathbf{s}, \mathbf{d}, \cdot) = \Theta_j(\mathbf{s}', \mathbf{d}, \cdot)$ almost everywhere for every $\mathbf{d}$ only when $\mathbf{s} = \mathbf{s}'$. Then the classes of histories with the same profile are exactly the level sets of $g_j$, and $g_j$ is a minimal sufficient statistic in the sense of Theorem~\ref{thm-quotient-minimality}.
\end{enumerate}

\end{proposition}The proof is in the Appendix. For the transition alone the classes are the $\sim_j$-classes. For the stacked primitive they are the exact-factoring classes, and $\mathbf{s}_j$ must then carry every dependence of the marginal utility on the history, as well as that of the marginals of the transition and the constraints. The proposition reduces the minimal state of $H$ to a statistic of its marginal; the modeler must name it.

\subsection{Discovery of Intermediate States}\label{sec:discovery}

To summarize early history for the transition, utility and constraints at once, stack the transition, the utility offset $U - U^0$ (the utility net of its value at a fixed reference late control) and the constraint functions into $H := (F, U - U^0, \zeta_1, \ldots, \zeta_p)$. Here $\zeta_1, \ldots, \zeta_p$ are the constraint functions that involve a late control. I require these functions to agree across histories, not only the feasible sets $\mathcal{C}_j(\mathbf{e}_j) = \{\mathbf{l}_j : \zeta_k(\mathbf{e}_j, \mathbf{l}_j) \geq 0 \text{ for all } k\}$ they cut out. This \textit{exact-factoring convention} refines $\approx_j$, strictly where distinct constraint profiles cut out the same feasible set, as when a constraint is slack. A constraint on the early variables alone belongs to an earlier stage and holds on every history with a feasible completion, so it leaves the stack at cut $j$. A constraint that depends on the early variables only through the transition, $\zeta_k = \tilde{\zeta}_k(F(\mathbf{e}_j, \mathbf{l}_j), \mathbf{l}_j)$ for a continuous $\tilde{\zeta}_k$, agrees across histories with the same transition profile. It cannot refine the classes, so I leave it out of the stack as well. Early variables whose gradients are free of the late controls in every primitive are \textit{aggregable} and collapse into aggregates, linear ones when $H$ is affine; the others \textit{pass through} whole (Algorithm~\ref{alg-discover-states} in the Appendix).

\begin{proposition}[Correctness of the Discovery Algorithm]\label{prop-discovery-correctness}Suppose each component of the stacked primitive $H = (F, U - U^0, \zeta_1, \ldots, \zeta_p)$ is affine in $\mathbf{e}_j$ with coefficients continuous in $\mathbf{l}_j$, where $U - U^0$ is the utility offset implementing the utility-gap clause of the joint relation $\approx_j$.\footnote{For a reference late control $\mathbf{l}^0 \in \mathcal{L}_j$, $U^0(\mathbf{e}_j) := U(\mathbf{e}_j, \mathbf{l}^0)$ and the offset is $U - U^0$. Offsets at two references differ by a term constant in $\mathbf{l}_j$, so nothing depends on the choice.} Suppose also that the set $\mathcal{E}_j^{\mathcal{C}}$ of histories with a feasible completion has nonempty interior. Then the statistic $g_j$ returned by Algorithm~\ref{alg-discover-states} is a sufficient statistic for the joint system on $\mathcal{E}_j^{\mathcal{C}}$. The aggregable coordinates are represented by a basis of the projection of the joint row space onto them, so the merged aggregates span it. If in addition the coefficient variation across the pass-through components is nondegenerate (differences of coefficient rows across late controls span the pass-through coordinate directions), then $g_j$ attains the minimal dimension among all continuous joint statistics sufficient on $\mathcal{E}_j^{\mathcal{C}}$, namely $\dim \mathcal{E}_j - \dim \bigcap_{\mathbf{l}_j} \ker \mathsf{A}_H(\mathbf{l}_j)$, where $\mathsf{A}_H(\mathbf{l}_j)$ is the stacked coefficient matrix, and its level sets on $\mathcal{E}_j$ are exactly the exact-factoring classes. Minimality is against the exact-factoring convention for constraints, which refines $\approx_j$.

\end{proposition}The proof, Corollary~\ref{cor-gradient-test} applied to $H$, is in the Appendix.

\subsection{The Decomposition Procedure}\label{sec:procedure}

The results of Section~\ref{sec:algebraic_characterization} and this section combine into a procedure. Given an ordering, the states discovery returns assemble, under the conditions of the theorem below, into a sequential decomposition whose stage objects are built from the primitives. Fix a reference plan $\mathbf{A}^0 = (\mathbf{a}^0_1, \ldots, \mathbf{a}^0_J)$. At cut $j$ the reference late control is the tail $(\mathbf{a}^0_j, \ldots, \mathbf{a}^0_J)$, with reference utility $U^0_j(\mathbf{e}_j) := U(\mathbf{e}_j, \mathbf{a}^0_j, \ldots, \mathbf{a}^0_J)$, and stage $j$'s utility is the change in reference utility when the stage's control replaces its reference value,
\begin{equation}
u_j(\mathbf{e}_j, \mathbf{a}_j) := U^0_{j+1}(\mathbf{e}_j, \mathbf{a}_j) - U^0_j(\mathbf{e}_j), \qquad U^0_{J+1} := U,
\end{equation}
with $U^0_1(\mathbf{X}) = U(\mathbf{X}, \mathbf{A}^0)$ added at the first stage, so that the stage utilities telescope to $U$. The stage set at a history with a feasible completion collects the stage controls after which the extended history still has a feasible completion.

\begin{theorem}[Discovered States Assemble into a Decomposition]\label{thm-assembly}Fix an ordered partition $(\mathbf{a}_1, \ldots, \mathbf{a}_J)$ of the controls. For $j = 2, \ldots, J$ let $g_j$ be a continuous map from $\mathcal{E}_j$ into a metric space whose level sets are exactly the exact-factoring classes at cut $j$, the classes of histories with the same profile $H_j(\mathbf{e}_j, \cdot)$ of the stacked primitive there. Suppose the feasible completions $\mathcal{C}_j$ are lower hemicontinuous on the histories that have one. Take the states $\mathbf{y}_0 = \mathbf{X}$, $\mathbf{y}_{j -1} = g_j(\mathbf{e}_j)$ and $\mathbf{y}_J = F$ on $\mathcal{Y}_0 = \mathcal{X}$, $\mathcal{Y}_{j -1} = g_j(\mathcal{E}_j)$ and $\mathcal{Y}_J = \mathcal{Y}$. With these stage utilities and sets, they form a sequential decomposition: every state is a continuous function of the previous state and the stage control, the stage utilities and stage sets depend on the history only through the entering state, and the stage sets are continuous and compact-valued on the reachable states. The states $\mathbf{y}_1, \ldots, \mathbf{y}_{J -1}$ identify exactly the exact-factoring classes.

\end{theorem}The proof is in the Appendix. Corollary~\ref{cor-feature-linear} and part 2 of Proposition~\ref{prop-marginal-inversion} deliver the level sets the first hypothesis asks for on all of $\mathcal{E}_j$, the exogenous states and discrete early controls included; part 2 does so for the household's state. Corollary~\ref{cor-gradient-test} and, under its nondegeneracy condition, Proposition~\ref{prop-discovery-correctness} deliver them on each slice where the discrete early variables, and any exogenous coordinate treated as a parameter, are fixed. But the slice states need not be the state on all of $\mathcal{E}_j$: with $F = e + l\varepsilon$ for an exogenous coordinate $\varepsilon$, the state on each slice of fixed $\varepsilon$ is $e$, but on all of $\mathcal{E}_j$ it is $(e, \varepsilon)$, which Corollary~\ref{cor-gradient-test} returns once $\varepsilon$ counts as an early variable. Joining the slices into one continuous map is therefore a separate check. Chaining needs the exact classes, because a state at the next cut whose level sets are finer than the classes need not be a function of the current state and the stage control. Eligibility thresholds violate lower hemicontinuity but fall under a closed-graph variant of the Equivalence Theorem. The Supplemental Appendix states that variant and treats slices that do not join.

The procedure takes the order of the controls as given, but discovery can choose it too. By the observation closing Section~\ref{sec:algebraic_characterization}, an order is a path through the sets $I$ of earlier controls that adds one control at a time. Write $d(I)$ for the covering dimension of the reachable minimal states after $I$, the exact-factoring classes, on the histories with a feasible completion, of the stacked primitive at the cut that places $I$ early, with $d(\varnothing) = n_0$, and $\mathcal{M}$ for the indices of the discrete controls. Even without lower hemicontinuity, the Chaining step of the proof of Theorem~\ref{thm-assembly} makes the state after $I \cup \{i\}$ a continuous function, the \textit{chaining map}, of the state after $I$ and $a_i$. The stage that chooses a scalar control $i$ after $I$ has $d(I) + 1$ continuous dimensions if $i$ is continuous and $d(I)$ if discrete.

\begin{proposition}[Choosing the Order]\label{prop-order-search}Let the controls be scalars. Say that (a) holds at a proper subset $I$ of them if the stacked primitive at its cut is affine in the continuous early variables on each slice where the discrete ones are fixed, each slice of the histories with a feasible completion having nonempty interior. Under (a), $d(I)$ is the largest dimension over the slices of the joint row space of the stacked coefficients (Corollary~\ref{cor-gradient-test}). Say that (b) holds at $I$ if the reachable states after $I$ embed in $\mathbb{R}^{d(I)}$ times a finite set and, for each control $i$ outside $I$, the chaining map, on the pairs of a state and $a_i$ with a feasible completion, is locally Lipschitz from the Euclidean metric to some metric inducing the topology of the states after $I \cup \{i\}$. Suppose for every proper $I$ and every $i$ outside $I$ with $I \cup \{i\}$ proper, (b) holds at $I$, or (a) holds at $I \cup \{i\}$ and, unless $I$ is empty, at $I$. Then:

\begin{enumerate}
\item For disjoint $I$ and $B$ whose union is a proper subset of the controls, $d(I \cup B) \leq d(I) + |B \setminus \mathcal{M}|$. Hence, for disjoint $I$ and $B$, splitting a stage that chooses the controls in $B$ after $I$ into scalar stages never raises the largest stage dimension.
\item Let $\overline{d}^*(\varnothing) = 0$ and $\overline{d}^*(I) = \min_{i \in I} \max\{\overline{d}^*(I \setminus \{i\}),\, d(I \setminus \{i\}) + [i \notin \mathcal{M}]\}$, the bracket one for a continuous control and zero for a discrete one. Then $\overline{d}^*$ of the full set is the smallest value, over all orders and all ordered partitions, of the largest stage dimension of a chain of minimal states, and backtracking the minimizers gives an order that attains it.
\end{enumerate}

\end{proposition}The Appendix gives the proof. A discrete control's values multiply the computational cost of its stage and of each later stage that carries it, without raising $\overline{d}$. Searching takes one discovery run for each nonempty proper subset of the controls, $2^{n_a} - 2$ runs, where screening every order cut by cut takes $n_a!\,(n_a - 1)$: for six controls, 62 against 3,600. The recursion over orders in part 2 holds in every model, with $d(I)$ possibly infinite. The best order yields a decomposition wherever the feasible completions at its cuts are lower hemicontinuous (Theorem~\ref{thm-assembly}). Condition (b) holds at every cut of the motivating example, whose states are Euclidean coordinates and whose chaining maps, such as $m = b + \theta n$, are polynomial. In the household model it holds at the empty set and all 62 cuts, with $d(I) = 3 + [\varsigma \in I] + [n \in I,\, z \notin I] + [c \in I,\, h' \notin I]$ (Supplemental Appendix): an early risky share, hours chosen before participation, and consumption chosen before the house each add a coordinate. The search singles out the 42 of 720 orders that place $\varsigma$ last, $h'$ before $c$, and $z$ before $n$ or right after it, that of Table~\ref{tab-hank-decomposition} among them. Only these reach four, the floor for a continuous stage on a three-dimensional state.

%% file: intro-two-asset.tex
\section{A Household Model with Housing, Portfolio Choice, and Participation}\label{sec:two-asset-example}

I apply the procedure of Section~\ref{sec:procedure} to a household problem in partial equilibrium. It extends the hours, consumption and portfolio choices of Section~\ref{sec:theoretical_foundations} with features the literature on heterogeneous households usually adds one at a time: a participation margin, a progressive tax on labor income, a house that carries an adjustment cost and serves as collateral, and an illiquid account. The household enters the period with an illiquid account $a$, liquid wealth $b$ (gross of the returns realized at the start of the period), and a house $h$. Its productivity $\varepsilon$, in efficiency units of labor, is an exogenous Markov state. It takes every price as given: the wage $w$ per efficiency unit, the house price $p_h$, and the returns $r^a$ on the illiquid account and $r^b$ on safe liquid assets. The household chooses participation $z \in \{0, 1\}$, hours $n$, a house $h'$, an illiquid deposit $a'$, consumption $c$, and a risky share $\varsigma$, with period utility
\begin{equation}
U = u(c, h') - \varphi \frac{n^{1+\nu}}{1+\nu} - \varpi z, \qquad u(c, h') = \frac{(c^{\alpha} h'^{1-\alpha})^{1-\sigma}}{1-\sigma},
\end{equation}
in which the house chosen this period delivers its services at once. Its budget is
\begin{equation}
\begin{aligned} c + \ell + a' + \Psi(a', a) + p_h h' + \Psi^{h}(h', h) &= b + (1+r^a)a + p_h(1-\delta)h \\ &\quad + w\varepsilon n - T(w\varepsilon n) + \iota(1-z), \end{aligned}
\end{equation}
where $\ell$ is final liquid assets, $\delta$ the depreciation rate of housing, $\varpi$ a fixed utility cost of working, $\iota$ a benefit paid to non-workers, and $T(y) = y - \tau_0 y^{1 -\tau_1}$ the tax on labor income $y$ under the progressive schedule of \citet{optimal_heathcote_2017}. For the illiquid account I use the adjustment cost $\Psi(a', a) = \frac{\chi_1}{2}\frac{((1+r^a)a - a')^2}{(1+r^a)a + \chi_0}$ with which \citet{sequence_auclert_2021} adapt the two-asset model of \citet{monetary_kaplan_2018} to discrete time. Housing bears the quadratic cost $\Psi^{h}(h', h) = \frac{\chi_h}{2}(h' - (1 -\delta)h)^2$. Workers supply at least $\underline{n}$ hours, $n \in [z\underline{n}, z\overline{n}]$. Requiring this minimum also keeps the after-tax marginal wage finite, because the marginal tax rate $T'(y)$ diverges to $-\infty$ as $y$ falls to zero. As in \citet{housing_kaplan_2020}, the household borrows against its house, $\ell \geq -\vartheta p_h h'$. It holds a share $\varsigma \in [0, 1]$ of any positive final liquid assets in the risky asset. Next period's liquid wealth is $b' = (1+r^b)\ell + (R' - 1 - r^b)\varsigma\max\{\ell, 0\}$, with the gross risky return $R'$ drawn by the law of motion after every decision of the period. The transition is
\begin{equation}
\begin{aligned} F &= \bigl(a',\; m - p_h h' - \Psi^{h}(h', h) - a' - \Psi(a', a) - c,\; h',\; \varsigma,\; \varepsilon\bigr) = (a', \ell, h', \varsigma, \varepsilon), \\ m &:= b + (1+r^a)a + p_h(1-\delta)h + w\varepsilon n - T(w\varepsilon n) + \iota(1-z). \end{aligned}
\end{equation}
Under this choice of controls, $(z, n, h', a', c, \varsigma)$ with $\ell$ the budget's residual, the model lies outside the triangular class (Supplemental Appendix): it has a discrete control, and its marginal adjustment costs depend on the pre-decision account and house, which the corrected conditions of \citet{multidimensional_corrigendum_2017} exclude.

\textit{Discovery.} This post-decision state is minimal, with the risky shares identified where $\ell \leq 0$: a nondegenerate return makes the law of next period's liquid wealth pin down $\ell$ and the risky total $\varsigma\max\{\ell, 0\}$. Productivity predicts its next value, with transition probabilities that differ across its values, so the post-decision state and every intermediate state carry $\varepsilon$, which I leave out of the notation below. On the finite grid on which productivity is usually discretized it is a discrete coordinate, outside the continuous dimensions counted below. With participation and hours ordered first, the early variables at the cut after hours are $(a, b, h, z, n)$ and the late controls are $(h', a', c, \varsigma)$. The rest of the period depends on liquid wealth, participation, and hours only through cash-on-hand $m$. Under the progressive tax, after-tax income is nonlinear in hours, yet each dollar of it enters the transition and the constraints as one more dollar of $m$. Both the illiquid account and the house pass through, each because its adjustment cost couples it to its own late control. Proposition~\ref{prop-marginal-inversion} turns these observations into the minimal state. Of the transition's marginals in the late controls, only those of $\ell$ in the house and the deposit vary with the history:
\begin{equation}
\frac{\partial \ell}{\partial h'} = -p_h - \chi_h\bigl(h' - (1-\delta)h\bigr), \qquad \frac{\partial \ell}{\partial a'} = -1 + \chi_1 \frac{(1+r^a)a - a'}{(1+r^a)a + \chi_0}.
\end{equation}
With $\chi_0, \chi_1, \chi_h > 0$, $\delta < 1$, $r^a > -1$ and nonnegative accounts $a, a'$, the first is strictly increasing in the old house $h$ and the second in the old account $a$. They depend on the early variables only through $(a, h)$, which they identify. The proposition's statistic adds the transition at a reference control, whose only history-dependent entry is $\ell$. At given $(a, h)$, that $\ell$ moves one for one with $m$. Part 2 of the proposition makes the $\sim_3$-classes the level sets of $(a, m, h)$. The joint state is the same, because the disutility of work is sunk once participation and hours are chosen, so the utility offset is free of the early variables. Like the collateral constraint, every constraint that involves both a late control and the early variables depends on the early variables only through $\ell$, so none can refine the classes (Section~\ref{sec:discovery}). The hours bounds drop out as well, because at this cut they do not involve a late control. This state has the smallest dimension of any (Supplemental Appendix). The same argument at the later cuts returns liquid assets $\ell^h = m - p_h h' - \Psi^h(h', h)$ after the housing choice, $\ell^a = \ell^h - a' - \Psi(a', a)$ after the illiquid choice, and final liquid assets $\ell = \ell^a - c$ after consumption, each carried with the positions chosen.

The cut after participation, by contrast, compresses nothing. Participation enters the bound on hours as well as cash-on-hand, so it passes through with $(a, b, h)$. With $z$ discrete, the hours stage searches over $n$ only for workers, with state $(a, b, h)$, four continuous dimensions in all; the hours of non-workers are fixed at zero. Placing the portfolio choice last keeps the risky share out of every stage state: the portfolio stage searches over $\varsigma$ on the positions $(a', \ell, h')$, while the four-dimensional post-decision state enters only the expectation over the return. Table~\ref{tab-hank-decomposition} lays out the resulting decomposition.

\begin{table}
\centering
\caption{Sequential decomposition of the household model}
\label{tab-hank-decomposition}
\begin{tabular}{llllll}
\toprule
Stage & Control & Input state & Stage utility & Transition & Output state \\
\midrule
1 & $z$ & $(a, b, h)$ & $-\varpi z$ & $z$ carried & $(a, b, h)$ per $z$ \\
2 & $n$ & $(a, b, h)$ per $z$ & $-\varphi n^{1+\nu}/(1+\nu)$ & $m$ & $(a, m, h)$ \\
3 & $h'$ & $(a, m, h)$ & $0$ & $\ell^h = m - p_h h' - \Psi^h(h', h)$ & $(a, \ell^h, h')$ \\
4 & $a'$ & $(a, \ell^h, h')$ & $0$ & $\ell^a = \ell^h - a' - \Psi(a', a)$ & $(a', \ell^a, h')$ \\
5 & $c$ & $(a', \ell^a, h')$ & $u(c, h')$ & $\ell = \ell^a - c$ & $(a', \ell, h')$ \\
6 & $\varsigma$ & $(a', \ell, h')$ & $0$ & $\varsigma$ carried & $(a', \ell, h', \varsigma)$ \\
\midrule
$W$ & & $(a', \ell, h', \varsigma)$ & & $R'$ realizes $b'$ & $(a', b', h')$ \\
\bottomrule
\end{tabular}
\end{table}
The discovered states assemble into a sequential decomposition by Theorem~\ref{thm-assembly}, whose lower-hemicontinuity hypothesis the Supplemental Appendix checks for this model. The table's stage utilities, the model's own terms, differ from the theorem's telescoped ones but also satisfy (S2). Each stage chooses one control on a three-dimensional state, so no stage has more than four continuous dimensions, against $3 + 5 = 8$ for the joint problem, which chooses all six controls (five continuous) on $(a, b, h)$. But the Supplemental Appendix's complexity theorem does not cover the drop from 8 to 4, because the stage sets fall outside its hypotheses.

At each stage, the marginal values of the intermediate value function are the shadow prices the household faces there: whether a household is hand-to-mouth is a property of the consumption stage's state, the hours condition is a Frisch labor supply, and participation kinks the value function (Supplemental Appendix).

%% file: intro-conclusion.tex
\section{Conclusion}\label{sec:conclusion}

The intermediate state is a property of the model's primitives, recoverable without ever solving the dynamic program. So is the computationally cheapest order of decisions, in both worked examples and wherever the primitives are affine. Cash-on-hand is the minimal sufficient statistic for the budget transition; the tests that return it find its analogues in models with no known aggregate.

Computing the minimal state reduces to linear algebra when the primitives have finite rank in the late controls. Beyond that class, Proposition~\ref{prop-marginal-inversion} gives the state globally when the modeler can name a statistic that the marginals identify. Otherwise only the local differential criterion of the Supplemental Appendix discovers the state, and only near points of constant rank. Rank can drop at the boundaries that matter economically (zero wealth, zero hours). Primitives beyond these classes still need a global test that lets a decomposition change as the binding constraints change.

%% file: intro-appendix.tex
\begin{appendix}
\section{The Discovery Algorithm}\label{app:discovery-algorithm}

\begin{observation}[Discovery of Intermediate States]\label{alg-discover-states}\textbf{Input:} Transition function $F: \mathcal{E}_j \times \mathcal{L}_j \to \mathcal{Y}$, utility function $U: \mathcal{E}_j \times \mathcal{L}_j \to \mathbb{R}$, and constraint functions $\zeta_k: \mathcal{E}_j \times \mathcal{L}_j \to \mathbb{R}$ for $k = 1, \ldots, p$ defining the feasible completions $\mathcal{C}_j(\mathbf{e}_j) = \{\mathbf{l}_j : \zeta_k(\mathbf{e}_j, \mathbf{l}_j) \geq 0,\ k = 1, \ldots, p\}$, each $C^1$ in the continuous early variables, with any discrete early control fixed on its slice as in Section~\ref{sec:algebraic_characterization}.

\textbf{Output:} An intermediate state function $g_j: \mathcal{E}_j \to \mathcal{Y}_{j -1}$, definite up to an invertible relabeling of its coordinates, such that $F$, the utility offset $U - U^0$ defined in step 1, and each $\zeta_k$ that involves a late control depend on $\mathbf{e}_j$ only through $g_j(\mathbf{e}_j)$, which determines the feasible completions on $\mathcal{E}_j^{\mathcal{C}}$.

\textbf{Procedure:}

\begin{enumerate}
\item \textit{Stack the primitives.} Fix a reference late control $\mathbf{l}^0$, set $U^0(\mathbf{e}_j) := U(\mathbf{e}_j, \mathbf{l}^0)$, form $H := (F, U - U^0, \zeta_1, \ldots, \zeta_p)$ over the constraint functions $\zeta_k$ that involve a late control, and compute the gradients $\nabla_{\mathbf{e}_j} H(\mathbf{e}_j, \mathbf{l}_j)$.
\item \textit{Classify each component} of $\mathbf{e}_j$ across \textit{all} primitives:

\begin{itemize}
\item $I_A = \left\{i : \partial H / \partial e_j^{(i)} \text{ is independent of } \mathbf{l}_j \text{ in every primitive}\right\}$ (aggregable)
\item $I_P$, the complement of $I_A$ (pass-through)
\end{itemize}

\item \textit{Construct the intermediate state.} Where $H$ is not affine in $\mathbf{e}_j$ and Corollary~\ref{cor-feature-linear} or Proposition~\ref{prop-marginal-inversion} applies to it, that result gives the whole state directly. Otherwise:

\begin{itemize}
\item Aggregable components collapse to the aggregate block $\mathbf{y}^{A}$, a map whose kernel on the $I_A$ block is the common kernel $\bigcap_{\mathbf{l}_j} \ker D_{\mathbf{e}_j} H$ restricted there. Outside the affine class, the local differential criterion of the Supplemental Appendix supplies its existence, dimension and minimality condition. When $H$ is affine in $\mathbf{e}_j$ the coefficients $\partial H^{(q)}/\partial e_j^{(i)}$ of each component $H^{(q)}$ are constants and the map is the linear aggregate $\sum_{i \in I_A} (\partial H^{(q)}/\partial e_j^{(i)})\, e_j^{(i)}$, with linearly dependent aggregates merged across primitives into a basis of the projection of the joint row space $\mathcal{R}$ onto the $I_A$ coordinates.
\item Pass-through components enter directly: $\mathbf{y}^{P} = (e_j^{(i)})_{i \in I_P}$.
\item The intermediate state is $\mathbf{y}_{j -1} = g_j(\mathbf{e}_j) = (\mathbf{y}^{A}, \mathbf{y}^{P})$.
\end{itemize}

\item \textit{Verify factorization.} Confirm that $F$, $U - U^0$ and each $\zeta_k$ involving a late control are functions of $(\mathbf{y}_{j -1}, \mathbf{l}_j)$ on all of $\mathcal{E}_j$. The check is needed for a state from the local differential criterion, which gives factorization only near each point; Corollary~\ref{cor-feature-linear} and Proposition~\ref{prop-marginal-inversion} prove it globally.
\end{enumerate}

\end{observation}\section{Proofs of the Main Results}\label{app:proofs_main}

\begin{proof}[Proof of the Equivalence Theorem]\label{proof-prop-sdp-hdp-equivalence}For each stage $j$, let $\Omega_j(\mathbf{y}_{j -1})$ be the maximum, over stage-wise feasible continuations $(\mathbf{a}_j, \ldots, \mathbf{a}_J)$ at reachable states, of the remaining utility $\sum_{k \geq j} u_k$ plus $W(\mathbf{y}_{J})$, and show $v_j = \Omega_j$ there by backward induction. By (S3), a reachable state has a stage-wise feasible continuation, and every $\mathbf{a}_j \in \gamma_j(\mathbf{y}_{j -1})$ leads to a reachable state, so both sides are evaluated only where they are defined. The base case is $\Omega_{J+1} = W = v_{J+1}$ at $\mathbf{y}_{J}$. For the step, the set of feasible continuations fibers over $\mathbf{a}_j$, and each later $\gamma_k$ depends on the history only through $\mathbf{y}_{k -1}$, so the joint maximum splits into an iterated one:
\begin{equation}
\Omega_j(\mathbf{y}_{j-1}) = \max_{\mathbf{a}_{j} \in \gamma_{j}(\mathbf{y}_{j-1})} \{ u_{j}(\mathbf{y}_{j-1}, \mathbf{a}_{j}) + \Omega_{j+1}(f_{j}(\mathbf{y}_{j-1}, \mathbf{a}_{j})) \},
\end{equation}
which is the recursion defining $v_j$. Hence $v_1 = \Omega_1$. Sequential decomposability (Definition~\ref{def-sequential-decomposability}) rewrites that maximization in HDP terms, invoking (S1) and (S2) only at plans that (S3) places in $\Gamma(\mathbf{X})$: for any feasible sequence, (S2) sums the stage utilities to $U(\mathbf{X}, \mathbf{A})$, (S1) identifies $\mathbf{y}_{J}$ with $F(\mathbf{X}, \mathbf{A})$, and (S3) identifies stage-wise feasibility with $\mathbf{A} \in \Gamma(\mathbf{X})$, so that
\begin{equation}
v_1(\mathbf{X}) = \max_{\mathbf{A} \in \Gamma(\mathbf{X})} \left\{ U(\mathbf{X}, \mathbf{A}) + W(F(\mathbf{X}, \mathbf{A})) \right\}.
\end{equation}
The right-hand side is the Bellman operator applied to $V$ in continuation-value form, and $V$ is its fixed point (Theorem~\ref{prop-hdp-solution}), hence $v_1 = V$; up to this step the argument used only that $W$ is bounded and continuous, so the staged recursion computes the joint maximum for any such terminal value. A plan whose stage choices solve their stage problems lies in $\Gamma(\mathbf{X})$ by (S3), as a joint maximizer does by definition. For $\mathbf{A} \in \Gamma(\mathbf{X})$ with states $\mathbf{y}_j$, all reachable by (S3), the stage slacks $\Delta_j := v_j(\mathbf{y}_{j -1}) - u_j(\mathbf{y}_{j -1}, \mathbf{a}_j) - v_{j+1}(\mathbf{y}_j)$ are nonnegative, because $\mathbf{a}_j \in \gamma_j(\mathbf{y}_{j -1})$, and vanish exactly when $\mathbf{a}_j$ solves its stage problem. Their sum telescopes, and (S1) and (S2) at the feasible plan give $\sum_j \Delta_j = v_1(\mathbf{X}) - U(\mathbf{X}, \mathbf{A}) - W(F(\mathbf{X}, \mathbf{A}))$. Since $v_1(\mathbf{X})$ is the joint maximum, $\mathbf{A}$ attains it exactly when every slack vanishes. Under the theorem's last hypothesis the objective is strictly concave on a convex set and has at most one maximizer.

\end{proof}\begin{proof}[Proof of the Necessity Theorem]\label{proof-thm-necessity}We prove that (i) implies (iii), that (iii) implies (ii), and that (ii) implies (i). The first implication is direct substitution: if $F = \tilde{F}(g(\cdot), \cdot)$, the objective depends on $(\mathbf{X}, \mathbf{a}_1)$ only through $g$, hence so does the maximum. The second is immediate because every $W \in \mathcal{W}$ and every penalty $-M\|\mathbf{a}_2 - \mathbf{a}_2^{\circ}\|^2$ on the compact $\mathcal{A}_2$ are bounded and continuous.

For (ii) implies (i), fix $\mathbf{a}_2^{\circ} \in \mathcal{A}_2$, take any $W \in \mathcal{W}$ with $|W| \leq \overline{W}$, and for $M > 0$ define
\begin{equation}
\Lambda_M(\mathbf{X}, \mathbf{a}_1) = \max_{\mathbf{a}_2 \in \mathcal{A}_2} \left\{ -M\|\mathbf{a}_2 - \mathbf{a}_2^{\circ}\|^2 + W(F(\mathbf{X}, \mathbf{a}_1, \mathbf{a}_2)) \right\}.
\end{equation}
By hypothesis (ii), $\Lambda_M$ depends on $(\mathbf{X}, \mathbf{a}_1)$ only through $g(\mathbf{X}, \mathbf{a}_1)$ for each $M$.

\textit{Localization and convergence.} Let $\mathbf{a}_2^*(M)$ denote a maximizer, which exists by Weierstrass since $\mathcal{A}_2$ is compact and the objective continuous; fix any selection. Suppressing $(\mathbf{X}, \mathbf{a}_1)$ in $F$, evaluating the objective at $\mathbf{a}_2^{\circ}$ gives $\Lambda_M \geq W(F(\mathbf{a}_2^{\circ}))$, and comparing this lower bound with the value at the maximizer yields $M\|\mathbf{a}_2^*(M) - \mathbf{a}_2^{\circ}\|^2 \leq 2\overline{W}$, so $\mathbf{a}_2^*(M) \to \mathbf{a}_2^{\circ}$ as $M \to \infty$. The bounds $W(F(\mathbf{a}_2^{\circ})) \leq \Lambda_M \leq W(F(\mathbf{a}_2^*(M)))$ and continuity of $W \circ F$ then give $\Lambda_M \to W(F(\mathbf{a}_2^{\circ}))$ pointwise.

\textit{Separation.} Write $\mathbf{e} = (\mathbf{X}, \mathbf{a}_1)$ and $\mathbf{e}' = (\mathbf{X}', \mathbf{a}'_1)$, and $F(\mathbf{e}, \mathbf{a}_2)$ for the transition at the history $\mathbf{e}$ and late control $\mathbf{a}_2$. Each $\Lambda_M$ factors through $g$, and so does their pointwise limit. Hence if $g(\mathbf{e}) = g(\mathbf{e}')$, then $W(F(\mathbf{e}, \mathbf{a}_2^{\circ})) = W(F(\mathbf{e}', \mathbf{a}_2^{\circ}))$ for every $W \in \mathcal{W}$. Because $\mathcal{W}$ separates points of $\mathcal{Y}$, it follows that $F(\mathbf{e}, \mathbf{a}_2^{\circ}) = F(\mathbf{e}', \mathbf{a}_2^{\circ})$. Since $\mathbf{a}_2^{\circ}$ was arbitrary, $F$ depends on $(\mathbf{X}, \mathbf{a}_1)$ only through $g$, and we may define $\tilde{F}(\mathbf{y}_1, \mathbf{a}_2)$ as the common value of $F$ on the $g$-fiber over $\mathbf{y}_1$. $\tilde{F}$ is continuous on the image of $(\mathbf{X}, \mathbf{a}_1, \mathbf{a}_2) \mapsto (g, \mathbf{a}_2)$, which contains every argument (i) evaluates, because that map is a continuous surjection from a compact domain onto a Hausdorff image, hence closed, so a function on the image is continuous whenever its composite with the map is. For the $J$-stage extension, run the argument at each cut $j$ with the early block $(\mathbf{X}, \mathbf{a}_1, \ldots, \mathbf{a}_{j -1})$ in place of $(\mathbf{X}, \mathbf{a}_1)$ and the compact block $(\mathbf{a}_j, \ldots, \mathbf{a}_J)$ in place of $\mathbf{a}_2$; it uses only compactness and continuity.

\end{proof}\begin{proof}[Proof of the Minimal-State Theorem]\label{proof-thm-quotient-minimality}\textit{Part 1.} Continuity of $\Phi_j$: $F$ is uniformly continuous on the compact $\mathcal{E}_j \times \mathcal{L}_j$, so $\mathbf{e}_j \mapsto F(\mathbf{e}_j, \cdot)$ is continuous into the supremum metric, and its image $\mathcal{Q}_j$ is compact. The equivalence $\mathbf{e}_j \sim_j \mathbf{e}'_j \iff \Phi_j(\mathbf{e}_j) = \Phi_j(\mathbf{e}'_j)$ restates the definition of $\sim_j$. Factorization: define $\tilde{F}(q, \mathbf{l}_j) := q(\mathbf{l}_j)$, the evaluation map, continuous on $C(\mathcal{L}_j, \mathcal{Y}) \times \mathcal{L}_j$; then $\tilde{F}(\Phi_j(\mathbf{e}_j), \mathbf{l}_j) = F(\mathbf{e}_j, \mathbf{l}_j)$.

\textit{Part 2.} Define $\tilde{\Phi}(g(\mathbf{e}_j)) := \Phi_j(\mathbf{e}_j)$. It is well-defined, as sufficiency of $g$ gives $\Phi_j(\mathbf{e}_j) = \Phi_j(\mathbf{e}'_j)$ whenever $g(\mathbf{e}_j) = g(\mathbf{e}'_j)$, and unique, as $g$ is onto $g(\mathcal{E}_j)$.

\textit{Part 3.} If two statistics each factor through the other, the connecting maps are mutually inverse bijections. When both statistics are continuous on the compact $\mathcal{E}_j$ with Hausdorff codomains, each image is compact, each statistic is a closed map onto its image (a continuous surjection from a compact space to a Hausdorff space is closed), so the connecting bijection is a homeomorphism.

\end{proof}\begin{proof}[Proof of the Chaining Lemma]\label{proof-lem-chaining}Well-definedness: suppose $\Phi_j(\mathbf{e}_j) = \Phi_j(\mathbf{e}'_j)$, that is, $F(\mathbf{e}_j, \mathbf{a}_j, \mathbf{l}_{j+1}) = F(\mathbf{e}'_j, \mathbf{a}_j, \mathbf{l}_{j+1})$ for all $(\mathbf{a}_j, \mathbf{l}_{j+1})$. Fixing the first coordinate at $\mathbf{a}_j$ and letting $\mathbf{l}_{j+1}$ range freely shows $\Phi_{j+1}(\mathbf{e}_j, \mathbf{a}_j) = \Phi_{j+1}(\mathbf{e}'_j, \mathbf{a}_j)$, so $\tilde{f}_j(\Phi_j(\mathbf{e}_j), \mathbf{a}_j) := \Phi_{j+1}(\mathbf{e}_j, \mathbf{a}_j)$ is well-defined. Continuity: $\Phi_j(\mathbf{e}_j)$ is the function $F(\mathbf{e}_j, \cdot, \cdot) \in C(\mathcal{A}_j \times \mathcal{L}_{j+1}, \mathcal{Y})$, and $\tilde{f}_j$ is the currying map $(q, \mathbf{a}_j) \mapsto q(\mathbf{a}_j, \cdot)$, which is continuous in the supremum metrics on compact domains. For (S1), induction on $j$ gives $\mathbf{y}_j = \Phi_{j+1}(\mathbf{X}, \mathbf{a}_1, \ldots, \mathbf{a}_j)$, ending in $\mathbf{y}_J = F$, and the folded $f_1$ is continuous as a composition. The fold matters: $\Phi_1$ sends distinct initial states $\mathbf{X}$ to one point of $\mathcal{Q}_1 = \Phi_1(\mathcal{X})$ whenever the initial state itself compresses, so a chain beginning at $\mathcal{Q}_1$ would miss (S1) as stated.

\end{proof}\begin{proof}[Proof of the Canonical Chain Corollary]\label{proof-cor-canonical-chain}The stage transitions $f_j, \ldots, f_J$ compute $F(\mathbf{e}_j, \mathbf{l}_j)$ from $\mathbf{y}_{j -1}(\mathbf{e}_j)$ and $\mathbf{l}_j$ by (S1), so $\mathbf{y}_{j -1}$ is a sufficient statistic at cut $j$, and part 2 of Theorem~\ref{thm-quotient-minimality} gives the unique $\tilde{\Phi}_j$. The chain of minimal sufficient statistics satisfies (S1) by Lemma~\ref{lem-chaining}. Two terminal chains each determine the other cut by cut, and part 3 of Theorem~\ref{thm-quotient-minimality} makes the connecting maps homeomorphisms, since the stage spaces are compact metric spaces.

\end{proof}\begin{proof}[Proof of Statement 3 of the Minimal Joint State Corollary]\label{proof-cor-joint-minimal-3}Fix $\mathbf{l}^{\circ}$ and $W$, and write $\Lambda_M(\mathbf{e}_j)$ for the maximum in statement 3 over $\mathcal{C}_j(\mathbf{e}_j)$, a nonempty closed subset of the compact $\mathcal{L}_j$; it exists because the objective is continuous. Let $\overline{B}$ bound $|U| + |W \circ F|$. If $\mathbf{l}^{\circ} \in \mathcal{C}_j(\mathbf{e}_j)$, the value at $\mathbf{l}^{\circ}$ gives $\Lambda_M \geq -\overline{B}$, so every maximizer $\mathbf{l}^*$ satisfies $M\|\mathbf{l}^* - \mathbf{l}^{\circ}\|^2 \leq 2\overline{B}$ and converges to $\mathbf{l}^{\circ}$, and continuity gives $\Lambda_M \to U(\mathbf{e}_j, \mathbf{l}^{\circ}) + W(F(\mathbf{e}_j, \mathbf{l}^{\circ}))$. If $\mathbf{l}^{\circ} \notin \mathcal{C}_j(\mathbf{e}_j)$, then $\Lambda_M \leq \overline{B} - M \operatorname{dist}(\mathbf{l}^{\circ}, \mathcal{C}_j(\mathbf{e}_j))^2$, which diverges to $-\infty$.

Suppose the condition holds and $g(\mathbf{e}_j) = g(\mathbf{e}'_j)$, so that $\Lambda_M(\mathbf{e}_j) - \Lambda_M(\mathbf{e}'_j) = \kappa(\mathbf{e}_j) - \kappa(\mathbf{e}'_j)$ for every $M$. Take $W = 0$. A point of $\mathcal{C}_j(\mathbf{e}_j)$ outside $\mathcal{C}_j(\mathbf{e}'_j)$ would send one side to a finite limit and the other to $-\infty$, so the feasible completions agree. For $\mathbf{l}^{\circ}$ in them the limits give $U(\mathbf{e}_j, \mathbf{l}^{\circ}) - U(\mathbf{e}'_j, \mathbf{l}^{\circ}) = \kappa(\mathbf{e}_j) - \kappa(\mathbf{e}'_j)$, a utility gap constant on the feasible completions. Subtracting this identity from the limits for $W \in \mathcal{W}$ leaves $W(F(\mathbf{e}_j, \mathbf{l}^{\circ})) = W(F(\mathbf{e}'_j, \mathbf{l}^{\circ}))$ for every $W$, and point separation gives $F(\mathbf{e}_j, \mathbf{l}^{\circ}) = F(\mathbf{e}'_j, \mathbf{l}^{\circ})$. Hence $\mathbf{e}_j \approx_j^{\mathcal{C}} \mathbf{e}'_j$.

Conversely, set $\kappa(\mathbf{e}_j) := \overline{U}(\mathbf{e}_j) = \max_{\mathbf{l} \in \mathcal{C}_j(\mathbf{e}_j)} U(\mathbf{e}_j, \mathbf{l})$, the best feasible utility. If $g(\mathbf{e}_j) = g(\mathbf{e}'_j)$, sufficiency makes $U(\mathbf{e}_j, \cdot) - U(\mathbf{e}'_j, \cdot)$ constant on the common feasible completions, where the transitions also agree. Hence $\kappa(\mathbf{e}_j) - \kappa(\mathbf{e}'_j)$ equals that constant. Each maximum net of $\kappa$ therefore takes the same value at both histories, so it depends on $\mathbf{e}_j$ only through $g(\mathbf{e}_j)$. None of $M$, $\mathbf{l}^{\circ}$ or $W$ enters $\kappa$.

\end{proof}\begin{proof}[Proof of the Affine Gradient Corollary]\label{proof-cor-gradient-test}\textit{Part 1.} Each row of $\mathsf{A}(\mathbf{l}_j)$ lies in $\mathcal{R} = \operatorname{rowspace}(\mathsf{C})$, so $\mathsf{A}(\mathbf{l}_j) = \mathsf{D}(\mathbf{l}_j) \mathsf{C}$ with $\mathsf{D}(\mathbf{l}_j) = \mathsf{A}(\mathbf{l}_j) \mathsf{C}^{+}$, continuous in $\mathbf{l}_j$, since $\mathsf{C}^{+}\mathsf{C}$ is the identity on $\mathcal{R}$.

\textit{Part 2.} For affine $F$, $F(\mathbf{e}_j, \mathbf{l}_j) = F(\mathbf{e}'_j, \mathbf{l}_j)$ for all $\mathbf{l}_j$ if and only if $\mathsf{A}(\mathbf{l}_j)(\mathbf{e}_j - \mathbf{e}'_j) = 0$ for all $\mathbf{l}_j$, that is, if and only if $\mathbf{e}_j - \mathbf{e}'_j$ is orthogonal to every row of every $\mathsf{A}(\mathbf{l}_j)$, hence to $\mathcal{R}$. The classes are therefore the cosets of $\mathcal{R}^{\perp}$ intersected with $\mathcal{E}_j$. Let $g$, a continuous map into a metric space, be defined and sufficient on $\mathcal{E}_j$ or on a subset of it with nonempty interior, so that every fiber of $g$ lies inside a single coset. Pick $\mathbf{e}^0$ in the interior of the domain of $g$ and let $\mathcal{N}$ be the intersection of the affine transversal $\mathbf{e}^0 + \mathcal{R}$ with a small open ball around $\mathbf{e}^0$, a relatively open $n_{j -1}$-dimensional disk inside that domain. Since $\mathcal{R} \cap \mathcal{R}^{\perp} = \{0\}$, the transversal meets each coset in at most one point, so $g$ is injective on $\mathcal{N}$. On a closed $n_{j -1}$-dimensional disk inside $\mathcal{N}$, $g$ is a continuous injection from a compact space, hence a homeomorphism onto its image. That image has covering dimension $n_{j -1}$ and lies in the image of $g$, a separable metric space as a continuous image of a subset of $\mathbb{R}^{\dim \mathbf{e}_j}$, whose covering dimension is therefore at least $n_{j -1}$ (\citet{dimension_hurewicz_1941}, Theorems III 1, IV 1 and V 8).

\end{proof}\begin{proof}[Proof of the Proposition on Late-Control Marginals]\label{proof-prop-marginal-inversion}\textit{Part 1.} Let $g_j(\mathbf{e}_j) = g_j(\mathbf{e}'_j)$, fix $\mathbf{d} \in \mathcal{D}_j$, and write $\boldsymbol{\Delta}(\mathbf{l}) := H(\mathbf{e}_j, \mathbf{d}, \mathbf{l}) - H(\mathbf{e}'_j, \mathbf{d}, \mathbf{l})$ on the slice $\mathcal{L}_j^{\mathbf{d}}$, whose interior $\mathcal{O}$ is open and convex. On $\mathcal{O}$, $\boldsymbol{\Delta}$ is locally Lipschitz. Let $\mathcal{V}$ be the set of points of $\mathcal{O}$ where $\boldsymbol{\Delta}$ fails to be differentiable with zero derivative. Both profiles are differentiable almost everywhere on $\mathcal{O}$ by Rademacher's theorem, and since $\mathbf{s}_j(\mathbf{e}_j) = \mathbf{s}_j(\mathbf{e}'_j)$, the hypothesis on the marginal makes their derivatives agree almost everywhere, so $\mathcal{V}$ is a null set. Take $\mathbf{l}, \mathbf{l}' \in \mathcal{O}$, set $\mathbf{v} := \mathbf{l}' - \mathbf{l}$, and choose $\rho > 0$ such that every segment from $\mathbf{l} + \mathbf{w}$ to $\mathbf{l}' + \mathbf{w}$ with $\mathbf{w}$ orthogonal to $\mathbf{v}$ and $\|\mathbf{w}\| \leq \rho$ lies in $\mathcal{O}$. Each such segment is a compact subset of $\mathcal{O}$, on which a locally Lipschitz map is Lipschitz, so $t \mapsto \boldsymbol{\Delta}(\mathbf{l} + \mathbf{w} + t\mathbf{v})$, $t \in [0, 1]$, is absolutely continuous. By Fubini's theorem in the coordinates $(\mathbf{w}, t)$, applied to a null Borel set containing $\mathcal{V}$, for almost every such $\mathbf{w}$ the points of the segment in $\mathcal{V}$ have parameters $t$ forming a null set. At every other point the derivative in $t$ is $\partial_{\mathbf{l}} \boldsymbol{\Delta}\,\mathbf{v} = 0$, so the function is constant on the segment and $\boldsymbol{\Delta}(\mathbf{l} + \mathbf{w}) = \boldsymbol{\Delta}(\mathbf{l}' + \mathbf{w})$. Such $\mathbf{w}$ come arbitrarily close to zero, and continuity gives $\boldsymbol{\Delta}(\mathbf{l}) = \boldsymbol{\Delta}(\mathbf{l}')$. Hence $\boldsymbol{\Delta}$ is constant on $\mathcal{O}$, and by continuity on its closure, the slice, since a box with nondegenerate intervals is the closure of its interior. The constant is $\boldsymbol{\Delta}(\mathbf{l}^{\mathbf{d}}) = 0$, because the reference evaluations are coordinates of $g_j$. As $\mathbf{d}$ was arbitrary, $H(\mathbf{e}_j, \cdot) = H(\mathbf{e}'_j, \cdot)$.

\textit{Part 2.} Let $H(\mathbf{e}_j, \cdot) = H(\mathbf{e}'_j, \cdot)$. The reference evaluations agree. On each slice the two profiles coincide, so their derivatives coincide wherever they exist, and the hypothesis on the marginal gives $\Theta_j(\mathbf{s}_j(\mathbf{e}_j), \mathbf{d}, \cdot) = \Theta_j(\mathbf{s}_j(\mathbf{e}'_j), \mathbf{d}, \cdot)$ almost everywhere for every $\mathbf{d}$. Identification gives $\mathbf{s}_j(\mathbf{e}_j) = \mathbf{s}_j(\mathbf{e}'_j)$, so $g_j(\mathbf{e}_j) = g_j(\mathbf{e}'_j)$. With part 1, the level sets of $g_j$ are exactly the classes. Being constant on each class, $g_j$ is a function of the profile map $\Phi_j$ of $H$. Every sufficient statistic determines $\Phi_j$ by part 2 of Theorem~\ref{thm-quotient-minimality} applied to $H$, and through $\Phi_j$ it determines $g_j$, which is the terminality property. Continuity of $g_j$ follows from that of $H$ and $\mathbf{s}_j$.

\end{proof}\begin{proof}[Proof of Correctness of the Discovery Algorithm]\label{proof-prop-discovery-correctness}The rows of the coefficient matrix $\mathsf{A}_H$ in the stacked primitive $H(\mathbf{e}_j, \mathbf{l}_j) = \mathsf{A}_H(\mathbf{l}_j)\mathbf{e}_j + \mathsf{B}_H(\mathbf{l}_j)$ collect the coefficients of $F$, the utility offset $U - U^0$, and the constraint functions $\zeta_k$. The gap $U(\mathbf{e}_j, \cdot) - U(\mathbf{e}'_j, \cdot)$ is constant in $\mathbf{l}_j$ exactly when the offsets $U - U^0$ agree at $\mathbf{e}_j$ and $\mathbf{e}'_j$, since their difference is the utility gap minus its value at $\mathbf{l}^0$. By Corollary~\ref{cor-gradient-test} applied to $H$, the exact-factoring classes are the cosets $\mathbf{e}_j + \mathcal{R}_H^{\perp}$, intersected with $\mathcal{E}_j$, of the orthogonal complement of the joint row space $\mathcal{R}_H = \operatorname{span} \bigcup_{\mathbf{l}_j} \operatorname{rowspace} \mathsf{A}_H(\mathbf{l}_j)$, and the minimal continuous statistic has dimension $n_{j -1} := \dim \mathcal{R}_H$.

\textit{Sufficiency of the algorithm's output.} Suppose $g_j(\mathbf{e}_j) = g_j(\mathbf{e}'_j)$, that is, the difference $\mathbf{u} := \mathbf{e}_j - \mathbf{e}'_j$ vanishes on every pass-through coordinate and is annihilated by every constructed aggregate. On the aggregable coordinates each row of $\mathsf{A}_H(\mathbf{l}_j)$ is constant in $\mathbf{l}_j$ (step 2) and a linear combination of the constructed aggregates (step 3), so it annihilates $\mathbf{u}$ there; on the pass-through coordinates $\mathbf{u}$ vanishes. Hence $\mathsf{A}_H(\mathbf{l}_j)\mathbf{u} = 0$ for all $\mathbf{l}_j$, so $H(\mathbf{e}_j, \cdot) = H(\mathbf{e}'_j, \cdot)$ and $g_j$ is sufficient.

\textit{Minimality under the stated nondegeneracy.} Write $d_A$ for the dimension of the projection of $\mathcal{R}_H$ onto the aggregable coordinates and $d_P$ for the dimension of $\mathcal{R}_H \cap \mathbb{R}^{I_P}$. Rank-nullity applied to that projection, restricted to $\mathcal{R}_H$, gives $n_{j -1} = d_A + d_P$, while the algorithm returns $\dim g_j = d_A + |I_P|$. Thus $\dim g_j \geq n_{j -1}$, with equality exactly when $d_P = |I_P|$, that is, when every pass-through unit vector lies in $\mathcal{R}_H$. This follows from the nondegeneracy hypothesis, since the differences $\mathsf{A}_H(\mathbf{l}_j) - \mathsf{A}_H(\mathbf{l}'_j)$, all in $\mathcal{R}_H$, are supported on the pass-through coordinates and span those directions as $(\mathbf{l}_j, \mathbf{l}'_j)$ vary. Hence $\dim g_j = n_{j -1}$, minimal on $\mathcal{E}_j^{\mathcal{C}}$ by part 2 of Corollary~\ref{cor-gradient-test} applied to $H$. The coordinates of $g_j$ are then linearly independent, so its kernel has dimension $\dim \mathcal{E}_j - n_{j -1}$, with $\dim \mathcal{E}_j$ the ambient dimension, which $\mathcal{E}_j$ fills because it has nonempty interior. That kernel lies in $\mathcal{R}_H^{\perp}$ by sufficiency, and the dimensions match, so it equals $\mathcal{R}_H^{\perp}$ and the level sets of $g_j$ are exactly the exact-factoring classes on $\mathcal{E}_j$.

\end{proof}\begin{proof}[Proof of the Assembly Theorem]\label{proof-thm-assembly}Write $g_1$ for the identity on $\mathcal{E}_1 = \mathcal{X}$ and $g_{J+1} := F$ on $\mathcal{E}_{J+1}$, and say that two histories at cut $j$ are \textit{equivalent} when their profiles $H_j(\mathbf{e}_j, \cdot)$ of the stacked primitive at that cut agree. Write $\mathcal{E}_j^{\mathcal{C}}$ for the histories with a feasible completion; it is compact, the projection of the compact graph of $\Gamma$. Leaving out constraints that read the early variables only through the transition does not change the classes: the transition's profile fixes theirs.

\textit{Chaining.} Let $2 \leq j \leq J$ and $g_j(\mathbf{e}_j) = g_j(\mathbf{e}'_j)$, so the two histories are equivalent. Fix $\mathbf{a}_j$. The transition profiles of $(\mathbf{e}_j, \mathbf{a}_j)$ and $(\mathbf{e}'_j, \mathbf{a}_j)$ at cut $j+1$ are slices of those at cut $j$, and so are the profiles of the constraints involving a control after stage $j$, all of which involve a late control at cut $j$. The utility offset at cut $j+1$ is the offset at cut $j$ minus its value at the reference completion, a term constant in the later controls. For $j < J$ the two extended histories are therefore equivalent at cut $j+1$, and $g_{j+1}$, constant on classes, takes the same value at both. At $j = J$ the transition rows give $F(\mathbf{e}_j, \mathbf{a}_j) = F(\mathbf{e}'_j, \mathbf{a}_j)$ directly. It follows that $f_j(g_j(\mathbf{e}_j), \mathbf{a}_j) := g_{j+1}(\mathbf{e}_j, \mathbf{a}_j)$ is well defined. At $j = 1$ the definition reads $f_1 = g_2$, since $g_1$ is the identity. By induction $\mathbf{y}_{j -1} = g_j(\mathbf{e}_j)$, so the generated pairs of stage $j$ form the image of $\mathcal{E}_j \times \mathcal{A}_j$ under $q_j: (\mathbf{e}_j, \mathbf{a}_j) \mapsto (g_j(\mathbf{e}_j), \mathbf{a}_j)$. The continuous surjection $q_j$ from a compact space is closed, so a function on its image is continuous when its composite with $q_j$ is (proof of the Necessity Theorem). Since $f_j \circ q_j = g_{j+1}$ is continuous, $f_j$ is continuous on the generated pairs. Composing the $f_j$ gives $\mathbf{y}_J = F$, which is (S1).

\textit{Utilities.} The stage utility of stage $j \geq 2$ is $U^0_{j+1}(\mathbf{e}_j, \mathbf{a}_j) - U^0_j(\mathbf{e}_j)$, and that of stage 1 is $U^0_2(\mathbf{X}, \mathbf{a}_1)$, so the sum telescopes to $U^0_{J+1} = U$, which is (S2). For $j \geq 2$ the stage utility is the utility offset at cut $j$, against the reference $(\mathbf{a}^0_j, \ldots, \mathbf{a}^0_J)$, evaluated at the completion $(\mathbf{a}_j, \mathbf{a}^0_{j+1}, \ldots, \mathbf{a}^0_J)$. Equivalent histories have offsets that agree against any reference, since offsets against two references differ by a term constant in the late controls, so the stage utility depends on $\mathbf{e}_j$ only through $g_j(\mathbf{e}_j)$. Since $q_j$ is a closed continuous surjection, the stage utility is continuous on the generated pairs, and bounded there by compactness.

\textit{Stage sets.} Let $\mathbf{e}_j, \mathbf{e}'_j \in \mathcal{E}_j^{\mathcal{C}}$ with $g_j(\mathbf{e}_j) = g_j(\mathbf{e}'_j)$. The constraints involving a late control agree as functions of the late controls, and those involving only early variables hold at both histories, since each has a feasible completion. Hence $\mathcal{C}_j(\mathbf{e}_j) = \mathcal{C}_j(\mathbf{e}'_j)$, and the stage set $\gamma_j(\mathbf{y}_{j -1})$, the projection of $\mathcal{C}_j(\mathbf{e}_j)$ onto $\mathcal{A}_j$ at any $\mathbf{e}_j \in \mathcal{E}_j^{\mathcal{C}}$ with $g_j(\mathbf{e}_j) = \mathbf{y}_{j -1}$, is well defined on the reachable states $\mathcal{P}_{j -1} = g_j(\mathcal{E}_j^{\mathcal{C}})$. It is nonempty, and compact as the projection of a closed subset of the compact $\Gamma(\mathbf{X})$. Since $\mathbf{e}_j$ itself represents its state, $\gamma_j(g_j(\mathbf{e}_j))$ is the projection of $\mathcal{C}_j(\mathbf{e}_j)$, so $(\mathbf{e}_j, \mathbf{a}_j)$ has a feasible completion exactly when $\mathbf{e}_j$ has one and $\mathbf{a}_j \in \gamma_j(g_j(\mathbf{e}_j))$, which is (S3). The graph of $\gamma_j$ is the image of the compact set $\mathcal{E}_{j+1}^{\mathcal{C}}$ under $(\mathbf{e}_j, \mathbf{a}_j) \mapsto (g_j(\mathbf{e}_j), \mathbf{a}_j)$, so it is closed, and $\gamma_j$ is upper hemicontinuous. For lower hemicontinuity at $\mathbf{y} \in \mathcal{P}_{j -1}$, let $\mathbf{y}_n \to \mathbf{y}$ in $\mathcal{P}_{j -1}$ and $\mathbf{a} \in \gamma_j(\mathbf{y})$, and choose $\mathbf{e}_n \in \mathcal{E}_j^{\mathcal{C}}$ with $g_j(\mathbf{e}_n) = \mathbf{y}_n$. Along any subsequence a further subsequence converges to some $\mathbf{e} \in \mathcal{E}_j^{\mathcal{C}}$ with $g_j(\mathbf{e}) = \mathbf{y}$, so $\gamma_j(\mathbf{y})$ is the projection of $\mathcal{C}_j(\mathbf{e})$. Lower hemicontinuity of $\mathcal{C}_j$ at $\mathbf{e}$, preserved by the projection, gives controls $\mathbf{a}_n \in \gamma_j(\mathbf{y}_n)$ converging to $\mathbf{a}$ along it. Every subsequence thus has a further subsequence along which the distance from $\mathbf{a}$ to $\gamma_j(\mathbf{y}_n)$ vanishes, so it vanishes along the whole sequence. At $j = 1$ the stage set is the projection of $\Gamma(\mathbf{X})$, continuous because $\Gamma$ is.

Hence the collection is a sequential decomposition. For $2 \leq j \leq J$ the level sets of $g_j$ are the exact-factoring classes by hypothesis, so $\mathbf{y}_{j -1}$ labels them one to one.

\end{proof}\begin{proof}[Proof of Choosing the Order]\label{proof-prop-order-search}Permuting early variables within a block is a homeomorphism commuting with the primitives, so $d(I)$ ignores their order.

\textit{Part 1.} It suffices to show $d(I \cup \{i\}) \leq d(I) + [i \notin \mathcal{M}]$ for each $i$ outside $I$ with $I \cup \{i\}$ proper, and add the controls of $B$ one at a time. The reachable states after $I \cup \{i\}$ are the image under the chaining map of the pairs with a feasible completion, which are compact as the image of the compact histories after $I \cup \{i\}$ with a feasible completion under $(\mathbf{e}, a_i) \mapsto (g_I(\mathbf{e}), a_i)$, with $g_I$ the minimal state after $I$.

\textit{Under (b) at $I$.} The pairs lie in $\mathbb{R}^{d(I) + [i \notin \mathcal{M}]}$ times a finite set, so their Hausdorff dimension is at most $d(I) + [i \notin \mathcal{M}]$. A locally Lipschitz map on a compact set is Lipschitz, so the reachable states after $I \cup \{i\}$, its image, have Hausdorff dimension no larger than the pairs in the metric of (b). The covering dimension of a separable metric space never exceeds its Hausdorff dimension, since a space of covering dimension $n$ has positive $n$-dimensional Hausdorff measure (\citet{dimension_hurewicz_1941}, Theorem VII 3). Hence $d(I \cup \{i\}) \leq d(I) + [i \notin \mathcal{M}]$.

\textit{Under (a) at $I \cup \{i\}$ and, unless $I$ is empty, at $I$.} On a slice the classes are the slice's intersections with the cosets of the orthogonal complement of its joint row space, of dimension $r$, so the slice's reachable states are homeomorphic to its orthogonal projection onto that row space, a compact set with nonempty interior in $\mathbb{R}^{r}$, of dimension $r$. These finitely many closed sets cover the reachable states, so by the sum theorem and monotonicity (\citet{dimension_hurewicz_1941}, Theorems III 2 and III 1) $d$ at a cut under (a) is the largest $r$ over its slices. For $I = \varnothing$ the bound follows, since $r$ is at most the number of continuous early variables after $\{i\}$. For nonempty $I$, a slice $\Sigma'$ after $I \cup \{i\}$ projects into the slice $\Sigma$ after $I$ that fixes the same discrete early variables of $I$'s block, and its continuous early variables are those of $\Sigma$ together with $a_i$ if it is continuous. A value of $a_i$ together with a completion $\mathbf{l}$ after $I \cup \{i\}$ forms a completion after $I$. The transition's coefficient on the early variables of $I$'s block after $I \cup \{i\}$ at $(a_i, \mathbf{l})$ is therefore its coefficient after $I$ at $(a_i, \mathbf{l})$, both being unique on the slices' interiors. The utility offset after $I \cup \{i\}$ at $(a_i, \mathbf{l})$ is the offset after $I$ at $(a_i, \mathbf{l})$ less that at $(a_i, \mathbf{l}^0)$, with $\mathbf{l}^0$ the reference completion after $I \cup \{i\}$. Each constraint in the stack after $I \cup \{i\}$ is either in the stack after $I$ or, having been left out there as a composition constraint $\tilde{\zeta}(F, \cdot)$, depends on the early variables of $I$'s block only through $F$. Hence a vector $\mathbf{v}$ of the common kernel of the stacked coefficients on $\Sigma$, extended by zero in a continuous $a_i$, lies in the common kernel on $\Sigma'$: for the left-out constraints because moving along $\mathbf{v}$ leaves $F$, and so $\tilde{\zeta}(F, \cdot)$, unchanged at every $(a_i, \mathbf{l})$. Joint row spaces are the orthogonal complements of common kernels, so the dimension on $\Sigma'$ exceeds that on $\Sigma$ by at most $[i \notin \mathcal{M}]$, as does the largest over slices.

Under either hypothesis, a stage that chooses the controls in $B$ after $I$ has $d(I) + |B \setminus \mathcal{M}|$ continuous dimensions. When that stage is split into scalar stages choosing $b_1, \ldots, b_k$ in turn, with $B_m = \{b_1, \ldots, b_m\}$, the stage choosing $b_m$ has $d(I \cup B_{m -1}) + [b_m \notin \mathcal{M}] \leq d(I) + |B_m \setminus \mathcal{M}| \leq d(I) + |B \setminus \mathcal{M}|$.

\textit{Part 2.} The stages of an order of the controls in $I$ that ends with $i$ are those of its restriction to $I \setminus \{i\}$ followed by the stage choosing $i$ after $I \setminus \{i\}$, with $d(I \setminus \{i\}) + [i \notin \mathcal{M}]$ continuous dimensions. Its largest stage dimension is the larger of the two, nondecreasing in the first, so minimizing over the orders of $I \setminus \{i\}$ and then over $i$ gives the recursion, and induction on the size of $I$ from $\overline{d}^*(\varnothing) = 0$ proves the claim. Recording a minimizing $i$ at each set and following these choices down from the full set gives an order whose largest stage dimension is $\overline{d}^*$ of the full set. The recursion needs neither (a) nor (b), and runs in the extended integers when some $d(I)$ is infinite. Under them, part 1 shows that the best order also attains the smallest largest stage over ordered partitions.

\end{proof}
\end{appendix}

%% file: intro-supplement.tex
\section{Supplemental Appendix}

\subsection{The Scope of the Theory}\label{app:scope}

The proof of the Equivalence Theorem evaluates only feasible plans and needs (S1) and (S2) only there, a weaker requirement that matters when early and late decisions interact only where the constraints forbid. With $a_1, a_2 \in [0, 1]$, $\Gamma = \{a_2 \leq 1/2\}$, $U = -a_1^2 - a_2^2$ and $F = a_1 \max\{0, a_2 - 1/2\}^2$, the transition vanishes on every feasible plan. On those plans a constant state decomposes, although the full transition needs $a_1$. Discovery, however, quantifies over all late controls and does not detect this compression.

\subsubsection{Recursive Preferences}

Recursive preferences fall within the theory. Under Epstein-Zin preferences \citep{substitution_epstein_1989}, the period objective is
\begin{equation}
\bigl[(1-\beta)U^{1-\rho} + \beta \mu^{1-\rho}\bigr]^{1/(1-\rho)}
\end{equation}
with $U > 0$, where $\mu$, the certainty equivalent of next period's value, depends on the period's decisions only through the post-decision state. It is a strictly increasing transformation of $\pi + \beta\mu^{1 -\rho}/(1 -\rho)$, with reward $\pi = (1 -\beta)U^{1 -\rho}/(1 -\rho)$ (logarithms at $\rho = 1$). Given the Epstein-Zin value function, whose existence is settled below, the within-period problem is the one studied here with $\pi$ in place of $U$ and $\beta\mu^{1 -\rho}/(1 -\rho)$ in place of $W$. The within-period conclusions of the Equivalence Theorem and the results of Sections 3 and 4 of the main text carry over. Risk aversion enters only $\mu$ and never changes the minimal state.

Stage additivity (S2) then applies to $\pi$, which can fail to be additive across stages where $U$ is. At the cut after hours in the main text's motivating model, part 2 of the main text's marginal-inversion proposition makes cash-on-hand alone the state exactly when the marginal reward of consumption is free of hours. Under expected utility with the separable form $u(c) - \varphi n^{1+\nu}/(1+\nu)$, the marginal $u'(c)$ is free of hours, so hours are sunk. Under Epstein-Zin, with $\sigma < 1$ and $\underline{c}^{1 -\sigma}/(1 -\sigma) > \varphi \overline{n}^{1+\nu}/(1+\nu)$ so that $U > 0$, the same form gives $\partial_c \pi = (1 -\beta)U^{ -\rho}u'(c)$; hours enter it through $U$ and stay in the state for every $\rho > 0$.

The aggregator need not satisfy Blackwell's conditions, so the contraction behind the main text's existence theorem does not apply. Existence and uniqueness follow instead from the argument of \citet{dynamic_ren_2020}, Section 3.2, which applies their Theorems 2.3 and 2.5 to this aggregator. On the compact domain here, with $U$ continuous and positive, Feller transitions, and $\Gamma$ continuous and compact-valued, the Bellman equation then has exactly one solution in the relevant class. That solution is the value function, defined as the supremum of lifetime utility over stationary policies; value function iteration converges to it, and an optimal policy exists. The argument requires $\rho > 0$ and risk aversion above $\rho$, the case of a preference for early resolution of uncertainty that they call empirically relevant, with neither equal to one; their Section 3.3 adds unit elasticity, $\rho = 1$ with risk aversion above one, through the risk-sensitive form after a logarithmic transformation. \citet{unique_marinacci_2010} and \citet{necessary_borovicka_2020} give conditions for existence and uniqueness of recursive utilities in more general settings.

\subsubsection{Boundedness, Compactness and the Law of Motion}

Utility bounded on one side only, such as power utility, falls under the positive and negative models of \citet{stochastic_bertsekas_1978}, Chapter 9, whose optimality equation (their Proposition 9.8) supplies the continuation value; the within-period results apply wherever that value is continuous and bounded on the post-decision states. Compactness of $\mathcal{X}$ and $\mathcal{A}$ gives Sections 3 and 4 of the main text their compact domains $\mathcal{E}_j$. When a formula for next period's state takes values outside $\mathcal{X}$ off the feasible plans, I follow it by the metric projection of the continuous coordinates onto their convex range. The projection is continuous and leaves every next-period state of a feasible plan unchanged. If the formula gives a Feller transition function and keeps the discrete coordinates in their finite set, so does its projection, since a bounded continuous function of the projected state is a bounded continuous function of the unprojected one.

\subsubsection{Stage Spaces and Eligibility Thresholds}

The Equivalence Theorem uses only compactness and continuity, so the spaces of minimal states of Section 3 of the main text (compact spaces of profiles) qualify as stage spaces. On the affine class the minimal sufficient statistic is a linear aggregate, so the stage space is a subset of $\mathbb{R}^{n_j}$ (the main text's affine gradient corollary); the complexity theorem of Section~\ref{sec:computational_implications} takes stage spaces in unit cubes.

Concave constraints with a strictly feasible point deliver the lower hemicontinuity that the assembly theorem of the main text asks of the feasible completions.

\begin{lemma}[Lower Hemicontinuity from Concave Constraints]\label{lem-slater-lhc}At cut $j$, let the late controls be continuous with a convex domain $\mathcal{L}_j$, and let each of the finitely many constraint functions $\zeta_k(\mathbf{e}_j, \mathbf{l}_j)$ be continuous and concave in $\mathbf{l}_j$. Then $\mathcal{C}_j$, restricted to the histories that have a feasible completion, is lower hemicontinuous at every history $\mathbf{e}^*_j$ with a Slater point $\mathbf{l}^{\circ}$, one at which $\zeta_k(\mathbf{e}^*_j, \mathbf{l}^{\circ}) > 0$ for every constraint function that involves a late control, and at every history with a unique feasible completion.

\end{lemma}\begin{proof}Lower hemicontinuity at $\mathbf{e}^*_j$ means that every open set $\mathcal{U}$ meeting $\mathcal{C}_j(\mathbf{e}^*_j)$ meets $\mathcal{C}_j(\mathbf{e}_j)$ at every history $\mathbf{e}_j$ near $\mathbf{e}^*_j$ with a feasible completion. At such histories the constraints on the early variables alone hold. Let $\mathbf{l}^* \in \mathcal{U} \cap \mathcal{C}_j(\mathbf{e}^*_j)$. At a history with a Slater point, $\mathbf{l}_t := (1 -t)\mathbf{l}^* + t\mathbf{l}^{\circ}$ lies in the convex $\mathcal{L}_j$, and concavity gives $\zeta_k(\mathbf{e}^*_j, \mathbf{l}_t) \geq (1 -t)\zeta_k(\mathbf{e}^*_j, \mathbf{l}^*) + t\zeta_k(\mathbf{e}^*_j, \mathbf{l}^{\circ}) > 0$ for $t \in (0, 1]$ and every constraint function involving a late control. Fix $t$ with $\mathbf{l}_t \in \mathcal{U}$. By continuity these finitely many functions stay positive at $(\mathbf{e}_j, \mathbf{l}_t)$ for $\mathbf{e}_j$ near $\mathbf{e}^*_j$, so $\mathbf{l}_t \in \mathcal{U} \cap \mathcal{C}_j(\mathbf{e}_j)$. At a history with the unique completion $\mathbf{l}^*$, the graph of $\mathcal{C}_j$ is closed and $\mathcal{L}_j$ compact, so $\mathcal{C}_j$ is upper hemicontinuous and $\mathcal{C}_j(\mathbf{e}_j) \subseteq \mathcal{U}$ for $\mathbf{e}_j$ near $\mathbf{e}^*_j$; being nonempty, it meets $\mathcal{U}$.

\end{proof}Quasi-concavity does not suffice: $\zeta(y, a) = \max\{0, a - 1\} - y$ on $a \in [0, 2]$ and $y \in [0, 1]$ is nondecreasing in $a$, hence quasi-concave, with the Slater point $a = 2$ at $y = 0$, yet its feasible set jumps from $[0, 2]$ at $y = 0$ to $[1 + y, 2]$. In the motivating model the constraints are affine in the late controls. Every history there without a Slater point has a unique completion in hours and consumption: before hours, those at which even full-time hours leave only the floor, $b + \theta\overline{n} = \underline{c}$; after hours, those with cash-on-hand at the floor. The risky-share set $[0, 1]$ is the same at every history, so the completions, a product of the two, are lower hemicontinuous there too. After consumption every history has the Slater point $\varsigma = 1/2$. In the household model the feasibility paragraph below proves the stage sets continuous directly.

Eligibility thresholds, such as a house purchase feasible only at or above a down payment, make $\Gamma$ upper but not lower hemicontinuous. Replacing continuity of $\Gamma$ and of the stage constraints by closed graphs, and $\mathcal{B}(\cdot)$ by bounded upper semicontinuous functions, the existence of the value function and the Equivalence Theorem continue to hold, with $V$, $W$ and the $v_j$ upper semicontinuous: each maximum is attained, is upper semicontinuous and has a measurable maximizing selection (\citet{stochastic_bertsekas_1978}, Proposition 7.33), and Feller transition functions and uniform limits preserve upper semicontinuity. The complexity theorem of Section~\ref{sec:computational_implications} does not cover thresholds, since it assumes stage sets with Lipschitz endpoints.

When the discovery results deliver the classes only on slices where exogenous coordinates of the state are fixed, and the slice maps do not join into one continuous map, the assembly construction still runs slice by slice. The Equivalence Theorem's proof, applied at one slice at a time, gives $v_1 = V$ on each. Each slice's problem then decomposes separately, but the stage objects may be discontinuous across slices, so the result need not be a sequential decomposition.

\subsection{Results and Proofs for Sections 3 and 4: Characterization and Discovery}\label{app:proofs_characterization}

The partial maps $F^{(i)}(\cdot, \mathbf{l}_j)$ of a primitive $F$ with values in $\mathbb{R}^m$, over its components $1 \leq i \leq m$ and the late controls $\mathbf{l}_j \in \mathcal{L}_j$, span a space of functions of the early variables, whose dimension I call the \textit{rank of $F$ in the late controls}. The rank is finite whenever $\mathcal{L}_j$ is a finite set of discrete controls, at most $m|\mathcal{L}_j|$, and whenever each component of $F$ is a finite sum of products of a function of $\mathbf{e}_j$ and a function of $\mathbf{l}_j$, as a transition linear in known functions of the early variables is.

\begin{corollary}[Primitives of Finite Rank in the Late Controls]\label{cor-finite-rank}Let $F: \mathcal{E}_j \times \mathcal{L}_j \to \mathbb{R}^m$ be continuous with finite rank $K$ in the late controls. Choose $K$ partial maps $c_k := F^{(i_k)}(\cdot, \mathbf{l}^{(k)})$ forming a basis and set $\mathbf{c} := (c_1, \ldots, c_K)$. Then $F = \mathsf{A}(\mathbf{l}_j)\, \mathbf{c}(\mathbf{e}_j)$ with $\mathsf{A}$ continuous, the $\sim_j$-classes are exactly the level sets of $\mathbf{c}$, and these $K$ outcome components therefore form a minimal sufficient statistic.

\end{corollary}\begin{proof}[Proof of the Finite-Rank Corollary]\label{proof-cor-finite-rank}The partial maps span the space, so a basis can be chosen among them. Evaluations at points of $\mathcal{E}_j$ separate the functions of the $K$-dimensional span, so $K$ of them, at points $\mathbf{e}^{(1)}, \ldots, \mathbf{e}^{(K)}$, are linearly independent on it and the matrix $\mathsf{M} := [c_k(\mathbf{e}^{(p)})]$ is invertible. Each partial map is a combination of the basis, $F^{(i)}(\cdot, \mathbf{l}_j) = \sum_k \mathsf{A}_{ik}(\mathbf{l}_j)\, c_k$, and evaluating at those points gives the row $\mathsf{A}_{i\cdot}(\mathbf{l}_j) = [F^{(i)}(\mathbf{e}^{(p)}, \mathbf{l}_j)]_p\, \mathsf{M}^{ -1}$, continuous in $\mathbf{l}_j$. If $\mathbf{c}(\mathbf{e}_j) = \mathbf{c}(\mathbf{e}'_j)$, the representation gives $F(\mathbf{e}_j, \cdot) = F(\mathbf{e}'_j, \cdot)$. Conversely, each $c_k$ is a partial map, so equal profiles give equal $\mathbf{c}$. Continuity enters only through that of $\mathsf{A}$, and each component of $H$ is continuous when the constraint functions $\zeta_k$ are, so the argument applies to $H$ verbatim.

\end{proof}Although the minimal state depends on the feasible set alone, whether the construction applies depends on how the constraints are written. Written as $m - c \geq 0$, the budget constraint ties cash-on-hand $m$ to consumption $c$ through two functions; written as $\arctan(m - c) \geq 0$, with the same feasible set, it ties them through infinitely many. I therefore write constraints in budget form.

Written this way, the constraints common in the models this paper cites fall into the classes discovery covers (Table~\ref{tab-constraints}). A \textit{composition} constraint depends on the early variables only through the transition, so it adds nothing to the stacked primitive of the main text's discovery algorithm; the others are bounds on a control, or have finite rank in the late controls, as when they are affine on each slice of a discrete control.

\begin{table}
\centering
\caption{Common constraints and their class}
\label{tab-constraints}
\begin{tabularx}{\textwidth}{@{}>{\raggedright\arraybackslash}X>{\raggedright\arraybackslash}X>{\raggedright\arraybackslash}X@{}}
\toprule
Constraint & Form & Class \\
\midrule
Borrowing limit \citep{uninsured_aiyagari_1994} & $\ell \geq \underline{\ell}$ & composition \\
Collateral \citep{housing_kaplan_2020, portfolio_cocco_2005} & $\ell \geq -\vartheta p_h h'$ & composition \\
Payment-to-income \citep{housing_kaplan_2020} & payment at most $\lambda$ times after-tax income & finite rank once hours are chosen \\
Risky-asset bound & $0 \leq \varsigma \leq 1$ & control bound \\
Hours with participation & $z\underline{n} \leq n \leq z\overline{n}$ & finite rank \\
Consumption floor & $c \geq \underline{c}$ & control bound \\
Adjust-or-not \citep{consumption_berger_2015} & $h' = (1-\delta)h$ unless the fixed cost is paid & finite rank \\
\bottomrule
\end{tabularx}
\end{table}
When the house left alone stays within its bounds, $(1 -\delta)h \in [\underline{h}, \overline{h}]$, and both branches are affordable at every state, adjust-or-not keeps $\Gamma$ continuous, as the union of the branch without adjustment and the adjustment set. An affordability threshold on the fixed cost is an eligibility threshold, treated above.

Discovery in the main text classifies early variables one at a time and can miss a combination the late controls scale together. For $F(\mathbf{e}, l) = l\,(e^{(1)} + e^{(2)})$ it reports no aggregate, yet $e^{(1)} + e^{(2)}$ suffices, a case the nondegeneracy condition of the main text's correctness proposition rules out.

The marginal-inversion proposition of the main text asks nothing of how $H$ depends on the early variables beyond continuity. Its derivative in the late controls may fail to exist on a null set, so the proposition admits kinks in the late controls, among them the brackets of a piecewise-linear tax on labor income when hours are chosen late. In the proposition's opening example, $\mathbf{s}_j$ is hours and the reference evaluation of the budget is cash-on-hand net of the reference consumption, so hours stay in the state beside cash-on-hand. With $u$ linear the marginal utility is constant and hours are sunk.

With only discrete late controls, the proposition identifies the statistic only when $\mathbf{s}_j$ is constant, and $g_j$ is then the primitive's whole profile (its values at the discrete controls). As for primitives linear in known features, the coordinates of $g_j$ can outnumber the dimension of the minimal state. For $F(e, l) = e + e\,l$ with a continuous late control $l \in [0, 1]$ and reference $l = 0$, the statistic is $(e, e)$, two coordinates for a one-dimensional minimal state.

The transitions that \citet{sequential_lujan_2026} solves by inverting a stage's first-order condition, with no stage utility, satisfy the hypotheses of the proposition, the stage's control $a$ being the late control and $\phi_1$, $\phi_2$ continuous. In the form $(\phi_1(e^{(1)}) + k\,a,\ \phi_2(e^{(2)}) + \psi(a))$ with $k \neq 0$ and $\psi$ continuously differentiable, the marginal $(k, \psi'(a))$ is free of the early variables, so $\mathbf{s}_j$ is constant and the state is the transition at a reference control. In the form $(\phi_1(e^{(1)}) + \kappa(a, e^{(2)}),\ a)$ with $\kappa$ continuously differentiable and $\kappa_a(a, \cdot)$ injective, the marginal $(\kappa_a(a, e^{(2)}), 1)$ identifies $e^{(2)}$, and the state is $(e^{(2)},\ \phi_1(e^{(1)}) + \kappa(a^0, e^{(2)}))$ at a reference control $a^0$. Inverting the stage's first-order condition requires exactly this identification.

\subsubsection{Construction versus Selection}

The discovery algorithm \textit{constructs} the intermediate state as an aggregate of the early variables. Selecting a subset of the original variables in place of the construction always yields a sufficient statistic (the full set qualifies), but the selection can be strictly larger than the constructed state, as the proposition below shows. Say that $F$ \textit{essentially depends} on an early variable $e^{(i)}$ if there exist two early configurations in $\mathcal{E}_j$, identical except in $e^{(i)}$, at which $F$ differs for some completion by late controls.

\begin{proposition}[Selection and the Minimal State]\label{thm-selection-impossibility}Fix a cut $j$ and write the early variables as $\mathbf{e}_j = (e^{(1)}, \ldots, e^{(d)})$. If $F$ essentially depends on every early variable, then \textit{no} proper subset of the original variables is a sufficient statistic, so every selection has cardinality $d$, even where a constructed state of dimension $n_{j -1} < d$ suffices.

\end{proposition}\begin{proof}A proper subset $I$ omits some $e^{(i)}$. Two configurations that differ only in $e^{(i)}$ and at which $F$ differs for some completion have the same projection onto $I$, and $F$ does not factor through that projection.

\end{proof}In the main text's household model, the transition essentially depends on the illiquid account, liquid wealth, the house, participation, and hours, so the smallest sufficient \textit{subset} at the cut after hours is all five early variables. The constructed state $(a, m, h)$ compresses them to three. Over several resources, a linear budget constraint compresses this way: the resources, each individually relevant, enter the future only through their weighted sum.

\subsubsection{The Local Differential Criterion}\label{app:differential_criterion}

For smooth primitives outside the affine and finite-rank classes, the algebraic characterization has an infinitesimal counterpart that converts the functional equation defining $\sim_j$ into a rank condition. The local differential criterion intersects the Jacobian kernels over late controls. In the following example a transition factors although its kernel moves. At a late control that switches the early variable's channel off, the kernel jumps, a case the classical ratio-of-partials form excludes by assuming non-vanishing gradients.

\begin{example}[A Factoring Transition Whose Kernel Moves]\label{ex-kernel-collapse}Let $F(e, l) = e^2 l$ with scalar early variable $e \in [ -1, 1]$ and scalar late control $l \in [ -1, 1]$. This transition factors through $g(e) = e^2$, so its minimal intermediate state is one-dimensional. But the kernel of $D_e F = 2el$ is $\{0\}$ when $e \neq 0$ and $l \neq 0$, and is all of $\mathbb{R}$ when $l = 0$: at $l = 0$ the late control annihilates the channel through which $e$ acts, and the kernel jumps. By contrast, the common kernel $\bigcap_l \ker D_e F(e, l) = \{0\}$ (for $e \neq 0$) reports the correct statistic dimension: $1 - 0 = 1$.

\end{example}\begin{definition}[Common Kernel and Regular Factorization]\label{def-common-kernel}Let $F$, a map from $\mathcal{E}_j \times \mathcal{L}_j$ to $\mathcal{Y}$, be $C^1$ in $\mathbf{e}_j$. The \textit{common kernel} at $\mathbf{e}_j$ is the subspace\footnote{At a boundary point of $\mathcal{E}_j$, $T_{\mathbf{e}_j}\mathcal{E}_j$ is the full ambient tangent space, so $\mathcal{K}_j(\mathbf{e}_j)$ is still a subspace; the criterion below concerns interior points only.}
\begin{equation}
\mathcal{K}_j(\mathbf{e}_j) := \bigcap_{\mathbf{l}_j \in \mathcal{L}_j} \ker D_{\mathbf{e}_j} F(\mathbf{e}_j, \mathbf{l}_j) \subseteq T_{\mathbf{e}_j}\mathcal{E}_j.
\end{equation}
A factorization $F = \tilde{F}(g(\cdot), \cdot)$ is \textit{regular} if $g$ is a $C^1$ submersion onto an $n_{j -1}$-dimensional manifold and $\ker D_{\mathbf{e}_j} g(\mathbf{e}_j) = \mathcal{K}_j(\mathbf{e}_j)$ for all $\mathbf{e}_j$.

\end{definition}\begin{proposition}[Local Differential Criterion for Sequential Factorizability]\label{prop-rank-characterization}Let the late-control set $\mathcal{L}_j$ be compact (a finite set of discrete controls qualifies), and let $F$ be $C^1$ in $\mathbf{e}_j$ and jointly continuous, with $\mathcal{E}_j$ a connected $C^1$ manifold (possibly with boundary or corners, as products of intervals are).

\begin{enumerate}
\item \textit{Lower bound.} If a $C^1$ submersion $g$ onto an $n_{j -1}$-dimensional manifold satisfies $F = \tilde{F}(g(\cdot), \cdot)$, then the kernel of $D_{\mathbf{e}_j} g(\mathbf{e}_j)$ lies in $\mathcal{K}_j(\mathbf{e}_j)$ at every interior $\mathbf{e}_j$; hence every submersive intermediate state satisfies
\begin{equation}
n_{j-1} \;\geq\; \dim \mathcal{E}_j - \dim \mathcal{K}_j(\mathbf{e}_j) \quad \text{at every interior } \mathbf{e}_j.
\end{equation}

\item \textit{Attainment and exactness.} Suppose $\mathcal{K}_j$ has constant dimension $k$ near an interior point of $\mathcal{E}_j$. Then the point has a neighborhood on which $F$ admits a regular factorization through a submersion $g$ of the minimal dimension $n_{j -1} = \dim \mathcal{E}_j - k$, and two points of that neighborhood are $\sim_j$-equivalent exactly when $g$ agrees on them. Locally, the minimal sufficient statistic of the main text's minimal-state theorem is therefore a $C^1$ submersion.
\end{enumerate}

\end{proposition}Constant rank can fail: in Example~\ref{ex-kernel-collapse} the common kernel jumps to all of $\mathbb{R}$ at $e = 0$, where no regular factorization covers the point, although the continuous statistic $e^2$ still does. Even where the rank is constant, the criterion establishes factorization only near each point, since a level set of the local state can have several components. Global factorization needs the direct check of step 4 of the discovery algorithm in the main text's appendix.

\begin{proof}[Proof of the Local Differential Criterion]\label{proof-prop-rank-characterization}\textit{Part 1.} Differentiability of $\tilde{F}$ is not assumed but follows: a $C^1$ submersion has local $C^1$ right inverses $s$ with $g \circ s = \mathrm{id}$, so $\tilde{F}(\cdot, \mathbf{l}_j) = F(s(\cdot), \mathbf{l}_j)$ is $C^1$ near each point of the image. By the chain rule, $D_{\mathbf{e}_j} F(\mathbf{e}_j, \mathbf{l}_j) = D_{\mathbf{y}} \tilde{F}(g(\mathbf{e}_j), \mathbf{l}_j) \cdot D_{\mathbf{e}_j} g(\mathbf{e}_j)$ for every $\mathbf{l}_j$, so $\ker D_{\mathbf{e}_j} g$ lies in every $\ker D_{\mathbf{e}_j} F(\cdot, \mathbf{l}_j)$ and hence in $\mathcal{K}_j(\mathbf{e}_j)$. Since $g$ is a submersion onto an $n_{j -1}$-manifold, $\dim \ker D_{\mathbf{e}_j} g = \dim \mathcal{E}_j - n_{j -1}$, which rearranges to the displayed inequality.

\textit{Part 2.} Fix an interior point. The first step, finite stacking, adds late controls $\mathbf{l}^{(1)}, \mathbf{l}^{(2)}, \ldots$ one at a time: while the intersection of their Jacobian kernels at the point strictly contains $\mathcal{K}_j$, some late control's kernel misses a vector of it, and adding that control lowers the dimension, so after at most $\dim \mathcal{E}_j$ steps the intersection equals $\mathcal{K}_j$. For the stacked map $S := (F(\cdot, \mathbf{l}^{(1)}), \ldots, F(\cdot, \mathbf{l}^{(N)}))$, $\ker DS$ contains $\mathcal{K}_j$, of dimension $k$, near the point, while lower semicontinuity of the rank of the continuous $DS$ keeps $\dim \ker DS$ at most its value $k$ at the point. Hence $\ker DS = \mathcal{K}_j$ and $S$ has constant rank $n_{j -1} = \dim \mathcal{E}_j - k$ near the point. In a chart, the rank theorem for $C^1$ maps (\citet{principles_rudin_1976}, Chapter 9), with its coordinate domain shrunk to a ball, yields, inside that neighborhood, a neighborhood $\mathcal{U}$ of the point, a $C^1$ submersion $g$ of $\mathcal{U}$ onto an open subset of $\mathbb{R}^{n_{j -1}}$ whose fibers are connected (each is the image of a ball cut by an affine subspace), and an injective immersion $\tilde{S}$ with $S = \tilde{S} \circ g$ on $\mathcal{U}$. Then $\ker D g = \ker D S = \mathcal{K}_j$, which lies in $\ker D_{\mathbf{e}_j} F(\cdot, \mathbf{l}_j)$ for every $\mathbf{l}_j$, so each $F(\cdot, \mathbf{l}_j)$ has zero derivative along the connected fibers of $g$ and is constant on them. Hence $F = \tilde{F}(g(\cdot), \cdot)$ on $\mathcal{U}$, a regular factorization, with $\tilde{F}$ continuous through local right inverses of $g$ as in part 1. Conversely, two $\sim_j$-equivalent points of $\mathcal{U}$ share their profile, so $S$ takes the same value at both, and injectivity of $\tilde{S}$ gives the same value of $g$. Two points of $\mathcal{U}$ are therefore $\sim_j$-equivalent exactly when $g$ agrees on them.

\end{proof}\subsubsection{Orders and Sufficient Conditions for Decomposability}

The minimal chains of different orders can differ, as the next example shows.

\begin{example}[Two Incomparable Orders]\label{ex-order-multiplicity}Let $F(X, a_1, a_2) = X + a_1 + a_2$ with additive utility $u_1(a_1) + u_2(a_2)$ and free constraints. Both orders decompose fully, the order $(a_1, a_2)$ with minimal sufficient statistic $g = X + a_1$ at its cut and the order $(a_2, a_1)$ with minimal sufficient statistic $g' = X + a_2$. Each is canonical for its own order, by the main text's canonical-chain corollary, and neither refines the other once both are pulled back to the space of $(X, a_1, a_2)$. Appending a second output equal to $a_1$ separates the two orders on computational cost. Under $(a_1, a_2)$, $a_1$ is chosen before the cut and enters the outcome directly; the cut must retain it alongside $X + a_1$ and compresses nothing. Under $(a_2, a_1)$, $a_1$ is chosen after the cut and $X + a_2$ suffices, so only that order is compressive.

\end{example}Once candidate state evolution functions are in hand, whether found by the discovery algorithm or guessed from economic intuition, we verify (S1)-(S3) by writing the transition, the utility and the constraints recursively through those states. The proposition below states these conditions, (SC1)-(SC3), with the stage maps continuous on stated spaces.

\begin{proposition}[Sufficient Conditions for Sequential Decomposability]\label{prop-sufficient-conditions}Suppose there exists an ordered partition $\mathbf{A} = (\mathbf{a}_1, \dots, \mathbf{a}_J)$ and functions $\{\hat{\mathbf{y}}_j\}_{j=0}^{J}$ with $\hat{\mathbf{y}}_j: \mathcal{X} \times \mathcal{A}_1 \times \cdots \times \mathcal{A}_j \to \mathcal{Y}_j$, where the compact sets $\mathcal{Y}_j \subseteq \mathbb{R}^{n_j}$ are specified in advance with $\mathcal{Y}_0 = \mathcal{X}$ and $\mathcal{Y}_J = \mathcal{Y}$, and each hatted map below has the stated $\mathcal{Y}_{j -1}$ and $\mathcal{Y}_j$ as its domain and codomain, such that, at every control vector $\mathbf{A} \in \mathcal{A}$, feasible or not:

(SC1) \textit{Explicit recursive structure of $F$.} $F(\mathbf{X}, \mathbf{A}) = \hat{\mathbf{y}}_{J}(\mathbf{X}, \mathbf{A})$, where for each $j = 1, \dots, J$:
\begin{equation}
\hat{\mathbf{y}}_{j}(\mathbf{X}, \mathbf{a}_1, \dots, \mathbf{a}_j) = \hat{f}_j(\hat{\mathbf{y}}_{j-1}(\mathbf{X}, \mathbf{a}_1, \dots, \mathbf{a}_{j-1}), \mathbf{a}_j)
\end{equation}
for some continuous functions $\hat{f}_j: \mathcal{Y}_{j -1} \times \mathcal{A}_j \to \mathcal{Y}_{j}$, with $\hat{\mathbf{y}}_0(\mathbf{X}) = \mathbf{X}$.

(SC2) \textit{Explicit stage additivity of $U$.}
\begin{equation}
U(\mathbf{X}, \mathbf{A}) = \sum_{j=1}^J \hat{u}_j(\hat{\mathbf{y}}_{j-1}(\mathbf{X}, \mathbf{a}_1, \dots, \mathbf{a}_{j-1}), \mathbf{a}_j)
\end{equation}
for some continuous bounded functions $\hat{u}_j: \mathcal{Y}_{j -1} \times \mathcal{A}_j \to \mathbb{R}$.

(SC3) \textit{Recursive constraint structure.} For each $j$, the history $(\mathbf{X}, \mathbf{a}_1, \dots, \mathbf{a}_j)$ completes to a plan in $\Gamma(\mathbf{X})$ if and only if $(\mathbf{X}, \mathbf{a}_1, \dots, \mathbf{a}_{j -1})$ does and
\begin{equation}
\mathbf{a}_j \in \hat{\gamma}_j(\hat{\mathbf{y}}_{j-1}(\mathbf{X}, \mathbf{a}_1, \dots, \mathbf{a}_{j-1})),
\end{equation}
for some correspondences $\hat{\gamma}_j$ into $\mathcal{A}_j$, continuous and compact-valued on the reachable states of the main text's stage-feasibility definition.

\textit{Then} the HDP is sequentially decomposable (main text) with the stated spaces $\mathcal{Y}_j$, $f_j$ and $u_j$ the restrictions of $\hat{f}_j$ and $\hat{u}_j$ to the generated pairs, and $\gamma_j = \hat{\gamma}_j$.

\end{proposition}Regularity of $F$, $U$, and $\Gamma$ does not imply the regularity demanded of $\hat{f}_j$, $\hat{u}_j$, and $\hat{\gamma}_j$, because the equations (SC1)-(SC3) determine the hatted objects only on the ranges of the $\hat{\mathbf{y}}_j$.

\begin{proof}[Proof of the Sufficient Conditions]\label{proof-prop-sufficient-conditions}Let $f_j$ and $u_j$ be the restrictions of $\hat{f}_j$ and $\hat{u}_j$ to the generated pairs, and $\gamma_j := \hat{\gamma}_j$ on the reachable states. By (SC1) and $\hat{\mathbf{y}}_0(\mathbf{X}) = \mathbf{X}$, the intermediate states these $f_j$ generate are $\mathbf{y}_j = \hat{\mathbf{y}}_j$, with values in the stated $\mathcal{Y}_j$, where $\mathcal{Y}_0 = \mathcal{X}$ and $\mathcal{Y}_J = \mathcal{Y}$. With this substitution, (SC1), (SC2) and (SC3) read as (S1), (S2) and (S3). Restrictions of continuous maps are continuous, the $u_j$ are bounded, and each $\gamma_j$ is continuous and compact-valued on the reachable states by hypothesis.

\end{proof}\subsection{The Complexity of Within-Period Optimization}\label{sec:computational_implications}

This section quantifies what decomposition saves, in the information-based complexity tradition of \citet{complexity_chow_1989} and \citet{randomization_rust_1997}. Fix the continuation value $W$ and consider computing the within-period value
\begin{equation}
\Omega(\mathbf{X}) = \max_{\mathbf{A} \in \Gamma(\mathbf{X})} \left\{ U(\mathbf{X}, \mathbf{A}) + W(F(\mathbf{X}, \mathbf{A})) \right\}
\end{equation}
to uniform accuracy $\eta$. In this section the state $\mathbf{X} = \mathbf{y}_0$ has $n_0$ continuous coordinates. The primitives are accessible only by evaluation: an oracle query returns $(U, F, W)$ at a requested point and, for the structured class, the stage primitives $(u_j, f_j)$ and the endpoints of the stage feasibility sets.

\begin{theorem}[Complexity Separation for Decomposable Problems]\label{thm-complexity-separation}Let $\mathcal{F}_{L}$ denote the class of within-period problems on $[0,1]^{n_0} \times [0,1]^{n_a}$, with $n_0, n_a \geq 1$ and $L \geq 1$, whose primitives and continuation value $W$ are $L$-Lipschitz. Let $\mathcal{G}_{L}$ denote the subclass of $\mathcal{F}_{L}$ with a known sequential decomposition whose stage spaces before the last are unit cubes, whose stage utilities and stage transitions are given as $L$-Lipschitz maps on the whole cubes that restrict to the decomposition's on the generated pairs, each transition mapping into the next cube or, at the last stage, into $\mathcal{Y}$, and whose stage feasibility sets are boxes $\gamma_j(\mathbf{y}_{j -1}) = \prod_k [\alpha_{jk}(\mathbf{y}_{j -1}), \beta_{jk}(\mathbf{y}_{j -1})] \subseteq [0,1]^{\dim \mathcal{A}_j}$ with $L$-Lipschitz endpoints, nonempty at every point of those cubes and not only at the reachable states, with restriction to the reachable states equal to the decomposition's stage sets. Write $\overline{d} := \max_{j} ( \dim \mathcal{Y}_{j -1} + \dim \mathcal{A}_j )$ for the largest stage dimension.

\begin{enumerate}
\item \textit{Lower bound without access to structure.} Any deterministic algorithm that returns an $\eta$-accurate within-period value function for every problem in $\mathcal{F}_L$ must make at least $(c L / \eta)^{n_0 + n_a}$ oracle queries in the worst case, for a constant $c > 0$ independent of dimension.
\item \textit{Upper bound with structure.} Stage-wise gridding with stage-wise optimization returns an $\eta$-accurate within-period value function for every problem in $\mathcal{G}_L$ using at most $J \cdot (C_{L,J} / \eta)^{\overline{d}}$ queries for $\eta$ below a constant of the same kind, where $C_{L,J}$ depends on $L$, the number of stages and $\overline{d}$, but not on $\eta$.
\end{enumerate}

\end{theorem}Knowing a decomposition and querying its stage primitives therefore lowers the worst-case exponent from $n_0 + n_a$ to at most $\overline{d}$. The hard instances of part 1 are themselves decomposable, so its lower bound is the cost of querying a problem without its stage primitives: for $n_a \geq 2$, with one control coordinate per stage, the bump of each instance depends on the early variables only through their distance in the maximum norm from the matching coordinates of its center, which gives $\overline{d} = n_0 + 1$. Because the gridded solver evaluates stage values at every knot of the cube, reachable or not, the class $\mathcal{G}_{L}$ requires stage sets on the whole cube.

\begin{lemma}[Lipschitz Propagation Through Stages]\label{lem-lipschitz-chain}Suppose each $u_j$ and $f_j$, defined on the whole stage cube with $f_j$ mapping into the next one ($\mathcal{Y}$ at the last stage), is Lipschitz in $(\mathbf{y}_{j -1}, \mathbf{a}_j)$ with constants $L_{u_j}$ and $L_{f_j}$, each $\gamma_j$ is nonempty, compact-valued and Lipschitz as a correspondence (the Hausdorff distance between $\gamma_j(\mathbf{y})$ and $\gamma_j(\mathbf{y}')$ is at most $L_{\gamma_j} \|\mathbf{y} - \mathbf{y}'\|$), and $W$ is Lipschitz with constant $L_W$. Define $L_{v_{J+1}} := L_W$ and, backward, $L_{v_j} := (L_{u_j} + L_{f_j} L_{v_{j+1}})(1 + L_{\gamma_j})$. Then each exact intermediate value function $v_j^*$ of the stage recursion, with $v_{J+1}^* = W$, is Lipschitz with constant $L_{v_j}$.

\end{lemma}\begin{proof}Backward induction; the base case is the hypothesis on $W$. Fix $\mathbf{y}, \mathbf{y}'$ and let $\mathbf{a}$ be optimal at $\mathbf{y}$ (the maximum is attained). The Hausdorff condition supplies $\mathbf{a}' \in \gamma_j(\mathbf{y}')$ with $\|\mathbf{a} - \mathbf{a}'\| \leq L_{\gamma_j}\|\mathbf{y} - \mathbf{y}'\|$, whence
\begin{equation}
v_j^*(\mathbf{y}) - v_j^*(\mathbf{y}') \leq \bigl(L_{u_j} + L_{f_j}L_{v_{j+1}}\bigr)\bigl(\|\mathbf{y} - \mathbf{y}'\| + \|\mathbf{a} - \mathbf{a}'\|\bigr) \leq L_{v_j} \|\mathbf{y} - \mathbf{y}'\|,
\end{equation}
using the inductive hypothesis on $v_{j+1}^*$ for the continuation term. Swap $\mathbf{y}$ and $\mathbf{y}'$ and combine.

\end{proof}\begin{proof}[Proof of the Complexity Separation Theorem]\label{proof-thm-complexity-separation}\textit{Part 1.} Take $F \equiv 0$, $W \equiv 0$, $\Gamma$ the full cube $[0,1]^{n_a}$, and let $U = b$ range over $L$-Lipschitz functions on the cube of dimension $d = n_0 + n_a$, so that $\Omega(\mathbf{X}) = \max_{\mathbf{A}} b(\mathbf{X}, \mathbf{A})$. A query reveals $b$ at one point at most. All other oracle answers are the same across the family. If $\eta > L/8$, then $(L/(16\eta))^{d} < 1$, and one query is needed: without any, the algorithm returns a fixed $\hat{\Omega}$, which at every $\mathbf{X}$ is more than $\eta$ from 0 or from $3\eta$, the within-period values of the constant utilities $U \equiv 0$ and $U \equiv 3\eta$. If $\eta \leq L/8$, partition the cube, up to a boundary strip, into $N_{\eta} = \lfloor L/(8\eta) \rfloor^{d}$ cells of side $8\eta/L$, and for the center $\mathbf{z}_i$ of cell $i$ let $b_i(\mathbf{z}) := \max\{0,\, 4\eta - L \|\mathbf{z} - \mathbf{z}_i\|_{\infty}\}$. Each $b_i$ is $L$-Lipschitz, peaks at $4\eta$ and vanishes outside the interior of cell $i$. Its within-period value $\Omega^{(i)}$ peaks at $4\eta$, while the zero problem has $\Omega^{(0)} \equiv 0$. A deterministic algorithm follows, on the all-zero answer sequence, a fixed query path. With fewer than $N_{\eta}$ queries the path places none in the interior of some cell $i'$, the algorithm returns the same $\hat{\Omega}$ on the zero problem and on $b_{i'}$, and it errs by at least $2\eta > \eta$ on one of them. Hence at least $N_{\eta} \geq (L/(16\eta))^{d}$ queries are needed.

\textit{Part 2.} A box with $L$-Lipschitz endpoints is Lipschitz in the Hausdorff metric with constant $L\sqrt{\dim \mathcal{A}_j} \leq L\sqrt{\overline{d}}$, and the boxes are nonempty and compact on the whole cubes, so Lemma~\ref{lem-lipschitz-chain} applies with $L_{u_j} = L_{f_j} = L_W = L$ and $L_{\gamma_j} = L\sqrt{\overline{d}}$, giving constants $L_{v_j}$ that depend only on $L$, $J$ and $\overline{d}$. For an integer $m \geq 1$ and $\varrho := 1/m$, grid each $\mathcal{Y}_{j -1}$ uniformly with $m + 1 \leq 2m$ points per coordinate, and at each state knot take as control grid the uniform grid with spacing at most $\varrho$ and at most $2m$ points per coordinate on the box $\gamma_j(\mathbf{y}_{j -1})$, which comes within $\varrho\sqrt{\overline{d}}$ of every feasible control. The computed update $\hat{\mathcal{T}}_j$, applied to a function $g$ on $\mathcal{Y}_j$, maximizes $u_j + g \circ f_j$ over that grid at each knot and interpolates the knot values piecewise multilinearly, and $\hat{v}_j := \hat{\mathcal{T}}_j \hat{v}_{j+1}$ from $\hat{v}_{J+1} = W$. Each stage-$j$ pass costs at most $3 (2m)^{\overline{d}}$ queries, counting one endpoint query per knot and the $W$ queries at stage $J$. Write $\mathcal{T}_j$ for the exact stage update, so that $v_j^* = \mathcal{T}_j v_{j+1}^*$. On the reachable states the extended maps and boxes agree with the decomposition's, and by (S3) every feasible stage choice leads to a reachable state, so backward induction gives $v_j^* = v_j$ there. Because every $\Gamma(\mathbf{X})$ is nonempty, the reachable states before stage 1 are all of $[0,1]^{n_0}$. Up to its step $v_1 = V$, the proof of the main text's Equivalence Theorem uses only that $W$ is bounded and continuous, so $v_1^* = v_1 = \Omega$. The computed update $\hat{\mathcal{T}}_j$ is non-expansive in the supremum norm, since a maximum over a fixed set is and multilinear interpolation forms convex combinations of knot values. Hence $\|\hat{v}_j - v_j^*\|_\infty \leq \|\hat{v}_{j+1} - v_{j+1}^*\|_\infty + \|\hat{\mathcal{T}}_j v_{j+1}^* - v_j^*\|_\infty$, and the last term involves only exact value functions. At a knot the objective $u_j + v_{j+1}^* \circ f_j$ is Lipschitz in $\mathbf{a}_j$ with constant $L_{u_j} + L_{f_j} L_{v_{j+1}}$, so the control grid loses at most that constant times $\varrho\sqrt{\overline{d}}$. Interpolating the $L_{v_j}$-Lipschitz $v_j^*$ across a cell of diameter at most $\varrho \sqrt{\overline{d}}$ loses at most $L_{v_j} \varrho \sqrt{\overline{d}}$. Telescoping from $\hat{v}_{J+1} = W$ gives $\|\hat{v}_1 - v_1^*\|_\infty \leq C \varrho$ with $C := \sqrt{\overline{d}} \sum_j (L_{u_j} + L_{f_j} L_{v_{j+1}} + L_{v_j})$. For $\eta \leq C$, take $m = \lceil C/\eta \rceil$, so that $C\varrho \leq \eta$ and $2m \leq 4C/\eta$; with $C_{L,J} := 12C$ the total query count is at most $3J(4C/\eta)^{\overline{d}} \leq J (C_{L,J}/\eta)^{\overline{d}}$, and the threshold $C$ depends only on $L$, $J$ and $\overline{d}$.

\end{proof}\subsection{The Examples: Feasibility, First-Order Conditions, and the Imai-Keane Model}\label{app:examples}

This section states the feasibility sets of the main text's two examples, the first-order conditions of the household model, and the sequential decomposition of the Imai-Keane human capital model.

\subsubsection{The Labor, Consumption and Portfolio Model}

\textit{Regularity.} The wage lies in $[\underline{\theta}, \overline{\theta}]$ with $\underline{\theta} > 0$, and the risky return $R'$ in $[0, \overline{R}]$ with $\overline{R} \geq R_f > 0$ and a nondegenerate law. I bound wealth so that the state space $\mathcal{X} = [0, \overline{b}] \times [\underline{\theta}, \overline{\theta}]$ is compact and feasible plans never leave it: saving lies in $[0, \overline{a}]$ with $\overline{a} = \overline{b}/\overline{R}$, so next period's wealth $(R_f + \varsigma(R' - R_f))a$ stays in $[0, \overline{b}]$ for every $\varsigma \in [0, 1]$. Hours are feasible when they leave resources for the consumption floor $\underline{c} > 0$, $\gamma_1(b, \theta) = [\max\{0, (\underline{c} - b)/\theta\}, \overline{n}]$, nonempty and continuous because $\underline{\theta}\,\overline{n} \geq \underline{c}$ and $\theta \geq \underline{\theta} > 0$. The consumption stage has $\gamma_2(m) = [\max\{\underline{c}, m - \overline{a}\}, m]$ at the reachable cash-on-hand $m \in [\underline{c}, \overline{b} + \overline{\theta}\,\overline{n}]$, and the portfolio stage $\gamma_3(a) = [0, 1]$. The controls range over $\mathcal{A} = [0, \overline{n}] \times [\underline{c}, \overline{b} + \overline{\theta}\,\overline{n}] \times [0, 1]$. For $\sigma \geq 1$ the floor keeps both $u_2$ and the period utility $U$ bounded, as the main text's regularity assumption and its definition of a sequential decomposition require; for $\sigma < 1$ the compact control space bounds them alone.

On feasible plans, where the projection is inactive, next period's wealth identifies the post-decision state: its variance $(\varsigma a)^2 \operatorname{Var} R'$ pins down the risky total $\varsigma a \geq 0$, and its mean $R_f a + (\mathbb{E}R' - R_f)\varsigma a$ then pins down $a$, and with it $\varsigma$ wherever $a > 0$. Hence $(a, \varsigma)$, with the risky shares identified at zero saving, is minimal.

\subsubsection{The Household Model with Housing, Portfolio Choice, and Participation}

\textit{Bounds.} Productivity, an exogenous Markov state taking finitely many values for the main text's dimension count, lies in $[\underline{\varepsilon}, \overline{\varepsilon}]$ with $\underline{\varepsilon} > 0$, and the risky return $R'$ in $[0, \overline{R}]$; the household takes the house price $p_h > 0$ as given. The controls satisfy $n \in [z\underline{n}, z\overline{n}]$, $h' \in [\underline{h}, \overline{h}]$ with $\underline{h} > 0$, $a' \in [0, \overline{a}]$, $c \in [\underline{c}, \overline{c}]$ with $\underline{c} > 0$ and $\overline{c}$ at least the largest attainable $\ell^a + \vartheta p_h\,\overline{h}$, $-\vartheta p_h h' \leq \ell \leq \overline{\ell}$, and $\varsigma \in [0, 1]$. The state space is $\mathcal{X} = [0, \overline{a}] \times [\underline{b}, \overline{b}] \times [\underline{h}, \overline{h}]$, times the finite grid of productivity, with $\underline{b} = -(1+r^b)\vartheta p_h\,\overline{h}$ and $\overline{b} = \max\{1+r^b, \overline{R}\}\,\overline{\ell}$. Next period's liquid wealth $b' = (1+r^b)\ell + (R' - 1 - r^b)\varsigma\max\{\ell, 0\}$ stays in $[\underline{b}, \overline{b}]$: when $\ell < 0$ the risky total is zero, so it is at least $-(1+r^b)\vartheta p_h h' \geq \underline{b}$, and when $\ell \geq 0$ it lies between 0 and $\max\{1+r^b, \overline{R}\}\, \ell$. Consumption and housing services lie in compact sets bounded away from zero, so period utility is bounded and continuous.

\textit{Feasibility.} Write $y(1) = \min_{\varepsilon}\{w\varepsilon\underline{n} - T(w\varepsilon\underline{n})\}$ and $y(0) = \iota$ for the least net income of a worker and of a non-worker. Since after-tax income $y - T(y)$ increases in $y$, a worker's least income comes from the least hours. I impose $\chi_1 \leq 2$ and, for both $z \in \{0, 1\}$,
\begin{equation}
y(z) > \underline{c} + \vartheta p_h\bigl((1+r^b)\overline{h} - \underline{h}\bigr) + p_h\,\delta\underline{h} + \tfrac{\chi_h}{2}(\delta\underline{h})^2.
\end{equation}
From every state and for either participation choice, the plan with $n = z\underline{n}$, $h' = \max\{(1 -\delta)h, \underline{h}\}$, $a' = \min\{(1+r^a)a, \overline{a}\}$, $\ell = -\vartheta p_h h'$, and $\varsigma = 0$ then leaves consumption strictly above $\underline{c}$. Its illiquid choice frees $(1+r^a)a - a' - \Psi(a', a) \geq (1 - \chi_1/2)((1+r^a)a - a') \geq 0$ resources, because $\Psi(a', a) \leq \frac{\chi_1}{2}((1+r^a)a - a')$ when $a' \leq (1+r^a)a$. Its housing choice costs at most $p_h\,\delta\underline{h} + \frac{\chi_h}{2}(\delta\underline{h})^2$, and the debt rolled over is at most $(1+r^b)\vartheta p_h\,\overline{h}$ against new borrowing of at least $\vartheta p_h\,\underline{h}$. Every hours choice in $[z\underline{n}, z\overline{n}]$ then leaves a feasible completion, and (SC3) holds with $\hat{\gamma}_1 = \{0, 1\}$ and $\hat{\gamma}_2 = [z\underline{n}, z\overline{n}]$. Each of the housing, illiquid, and consumption stage sets collects the control values that leave a feasible completion, namely those at which the stage's slack (the resources passed on plus the borrowing capacity they support), a concave function of the stage's control, weakly exceeds a bound. At reachable stage states where the largest slack strictly exceeds the bound, the set is a nonempty interval with continuous endpoints. Where the largest slack just meets the bound, the set shrinks to the slack's maximizer. Strict concavity of the slack in the housing and illiquid controls and its strict decrease in consumption make that maximizer unique and continuous in the stage state. Given these sets and the portfolio set $[0, 1]$, the stage correspondences are continuous, nonempty, and compact-valued on the reachable states. So are $\Gamma$ and each set of feasible completions $\mathcal{C}_j$, their sequential compositions along continuous transitions. This verifies the lower-hemicontinuity hypothesis of the main text's assembly theorem.

\textit{Minimal states at every cut.} At the cut that places a set $I$ of controls early, the minimal state consists of productivity, final liquid assets $\ell$ at a reference completion, $h'$ if the house is early and $h$ otherwise, $a'$ if the deposit is early and $a$ otherwise, $\varsigma$ if it is early, $n$ if hours precede participation, $c$ if consumption precedes the house, and $z$ if participation precedes hours. In the stacked primitive's profile, $h$ shows through the cross term of $\Psi^h$ with the late $h'$, $a$ through $\partial \ell/\partial a'$, $n$ and $z$ through the hours bounds, $c$ through the factor $c^{\alpha(1 -\sigma)}$ on the late house's utility ($0 < \alpha < 1$, $\sigma \neq 1$), and early positions as coordinates of $F$. Histories that agree on these variables have equal profiles, since after-tax income and the participation terms are additive in $n$ and $z$. From liquid wealth just below $\overline{b}$, moving the feasibility plan inward gives a plan with every inequality strict, $\ell > 0$ and $0 < \varsigma < 1$, near which each coordinate has a variable of its own and the reachable states fill an open set of $\mathbb{R}^{d(I)}$ times a finite set, with $d(I) = 3 + [\varsigma \in I] + [n \in I,\, z \notin I] + [c \in I,\, h' \notin I]$. A continuous sufficient statistic is injective on a $d(I)$-dimensional disk there, so invariance of domain (proof of part 2 of the main text's affine gradient corollary) makes this state minimal. Each chaining map adds to $\ell$ a term in the new control and the carried positions, smooth except after-tax income, Lipschitz on the feasible hours $\{0\} \cup [\underline{n}, \overline{n}]$. Condition (b) of the order-search proposition thus holds at the empty set and all 62 cuts. Its recursion gives four, the floor since every $d(I) \geq 3$, attained by orders placing $\varsigma$ last, $h'$ before $c$, and $z$ before $n$ or right after it.

\textit{Complexity.} The hours, consumption and portfolio stage sets are intervals with Lipschitz endpoints. The housing and illiquid sets are too on stage states where their slack keeps a uniform margin: by concavity the slack's derivative in the control is bounded away from zero at an endpoint there, which makes the endpoint Lipschitz in the stage state. Where a housing or illiquid set shrinks to a point, however, its endpoints move like a square root. The housing, illiquid and consumption stage sets are also defined only at reachable states. The complexity theorem of Section~\ref{sec:computational_implications}, a statement about whole problems with Lipschitz box sets on whole unit cubes, therefore does not apply to the model as stated. For this model, no bound proved here covers the drop from eight joint dimensions to four per stage.

Where the stage values are differentiable and the choices interior, with the consumption floor and the bounds on the positions slack, each stage value reads as follows.

The consumption stage value $v_5(a', \ell^a, h')$ has marginal value $\partial_{\ell^a} v_5 = u_c(c^*, h')$, the marginal value of liquid assets, whether or not the collateral constraint binds. The marginal propensity to consume is the derivative of $c^*(a', \ell^a, h')$ in liquid assets. It equals one where the collateral constraint binds. Whether a household is hand-to-mouth is therefore a property of the stage-5 state. A household at its collateral constraint with a large illiquid account or home equity is the wealthy hand-to-mouth household of \citet{model_kaplan_2014}.

A marginal dollar of each stage's resource is worth the same at the next stage, so $\partial_m v_3 = \partial_{\ell^h} v_4 = \partial_{\ell^a} v_5 = u_c$. For a worker at interior hours, the hours condition
\begin{equation}
\varphi n^{\nu} = w\varepsilon \bigl(1 - T'(w\varepsilon n)\bigr)\, \partial_m v_3(a, m, h)
\end{equation}
is then the Frisch labor supply of \citet{empirical_macurdy_1981}, with the marginal utility of wealth given by the stage-3 marginal value. Hours depend on wealth only through $(a, m, h)$ and on progressivity only through the after-tax marginal wage.

The portfolio stage value $v_6(a', \ell, h')$ is the end-of-period value of the household's positions. Its risky share $\varsigma^*(a', \ell, h')$ depends on the end-of-period positions alone. Whether the household reached $\ell$ by working longer, consuming less, or starting from a larger house does not enter. The house and the illiquid account enter the portfolio rule as positions held alongside $\ell$, the position channel through which housing crowds out stockholding in \citet{portfolio_cocco_2005}, whose house-price risk this model omits. Productivity enters through the continuation value, the channel through which future labor income acts as a safe asset in \citet{consumption_cocco_2005}. I draw $R'$ independently of productivity. The main text's law of motion also allows the covariance of returns with productivity that \citet{consumption_cocco_2005} study; the post-decision state is the same under it. The marginal value $\partial_{\ell} v_6$ is the marginal value of saving. Away from $\ell = 0$, where the risky total $\varsigma\max\{\ell, 0\}$ kinks $v_6$, and with the collateral constraint slack, the Euler equation $u_c(c, h') = \partial_{\ell} v_6$ holds.

At the illiquid stage, the first-order condition $\partial_{a'} v_5 = (1 + \partial \Psi/\partial a')\, \partial_{\ell^a} v_5$ states that a dollar in the illiquid account is worth $1 + \partial \Psi/\partial a'$ liquid dollars. The marginal adjustment cost is the wedge between illiquid and liquid dollars. It depends only on the deposit $a'$ and the pre-decision account $a$. At the housing stage, $\partial_{h'} v_4 = (p_h + \partial \Psi^h/\partial h')\, \partial_{\ell^h} v_4$ equates the marginal value of housing to its marginal cost in liquid units, the house price plus the marginal adjustment cost. The envelope condition gives $\partial_h v_3 = \chi_h(1 -\delta)(h' - (1 -\delta)h)\, \partial_{\ell^h} v_4$. Cash-on-hand already counts the old house at its resale value, so at fixed $m$ the old house matters only through the cost of moving away from it.

At the participation stage, $v_1(a, b, h) = \max\{v_2(a, b, h, 0), v_2(a, b, h, 1) - \varpi\}$ is the upper envelope of the two hours-stage values, and the household works exactly when $v_2(a, b, h, 1) - \varpi \geq v_2(a, b, h, 0)$. Where that difference increases with productivity, it defines a reservation productivity. Where the two values cross, $v_1$ has a kink that reaches every stage of the preceding period through the continuation value, the non-concavity that DC-EGM \citep{endogenous_iskhakov_2017} handles.

\textit{First-order conditions.} In the unknowns $(n, h', a', c, \varsigma)$, with $\ell$ the budget's residual, the first-order conditions are coupled as written. The marginal utility of consumption depends on $c$ and $h'$ and enters the conditions for hours, housing, and the illiquid deposit, and the continuation value's arguments move with every control. The model also fails the triangularity conditions of \citet{multidimensional_iskhakov_2015} as corrected in \citet{multidimensional_corrigendum_2017}. Corrected Assumption A1 requires each derivative of a post-decision state with respect to a control to depend on the post-decision states and the exogenous state alone. Here $\partial \ell/\partial a' = -1 - \Psi_{a'}(a', a)$ with $\Psi_{a'}(a', a) = -\chi_1((1+r^a)a - a')/((1+r^a)a + \chi_0)$, and $\partial \ell/\partial h' = -p_h - \chi_h(h' - (1 -\delta)h)$. For $\chi_1, \chi_h > 0$ both vary with the pre-decision account $a$ or house $h$ at fixed post-decision states $(a', \ell, h', \varsigma)$. With Cobb-Douglas services the utility Hessian in $(c, h')$ is a full block. No permutation makes it triangular, so Assumption A4a fails as well. Participation is also a discrete choice.

\subsubsection{The Imai-Keane Human Capital Model}

\citet{multidimensional_iskhakov_2015} presented the model of \citet{intertemporal_imai_2004} as satisfying their triangularity conditions. Their corrigendum \citep{multidimensional_corrigendum_2017} corrected Assumption A1 and stated that ``with the above correction, the example in Section 4 is invalid.'' Yet the model is sequentially decomposable.

\textit{HDP Problem Setup.} In the Imai-Keane model, agents choose consumption $c_t$ and hours $n_t$ over a life cycle indexed by $t$ to maximize expected discounted utility:

\begin{equation}
V_{t}(A_t, K_t, \varepsilon_{t}) = \max_{c_t \ge 0, n_t \ge 0} \{\Upsilon(t) c_t^{\alpha_1}/\alpha_1 - \varphi \varepsilon_{t} n_t^{\alpha_2}/\alpha_2 + \beta \mathbb{E}_t [ V_{t+1} (A_{t+1}, K_{t+1}, \varepsilon_{t+1} ) ]\}.
\end{equation}

The state vector $\mathbf{X}_t = (A_t, K_t, \varepsilon_t)$ comprises assets $A_t$, human capital $K_t$ and a scalar labor disutility shock $\varepsilon_{t}$, a Markov process. Agents choose consumption $c_t$ and hours worked $n_t$, the control vector $\mathbf{A}_t = (c_t, n_t)$. The utility function is $U = \Upsilon(t) c_t^{\alpha_1}/\alpha_1 - \varphi \varepsilon_{t} n_t^{\alpha_2}/\alpha_2$, with $\alpha_1<1$, $\alpha_2>1$, disutility weight $\varphi > 0$, and an age-dependent shifter $\Upsilon(t)$. End-of-period assets follow the post-decision state transition $F(\cdot)$, $A_{t+1} = (1 +r)(A_t + R_k K_t n_t - c_t)$, where the wage is $R_k K_t$. Next period's human capital is $K_{t+1} = K^{d}_{t+1} \xi_{t+1}$, where $\xi_{t+1}$ is a transitory human capital shock realized after the agent makes the period $t$ decision, and $K^{d}_{t+1} = k_0 + \delta K_t + \omega(K_t, n_t, t)$ is the deterministic part with learning-by-doing:
\begin{equation}
\omega(K_t, n_t, t) = A_0(1 + A_1(t-19))(B_1 + K_t)\left[(n_t + d_1)^{\alpha_3} - B_2(n_t + d_1) \right].
\end{equation}
I rename symbols of the original that this paper uses for other objects: the accumulation exponent $\alpha$ becomes $\alpha_3$, the accumulation function $g$ becomes $\omega$ ($g$ is the intermediate-state statistic), the age shifter $A(t)$ becomes $\Upsilon(t)$, the disutility weight $b$ becomes $\varphi$, hours $h_t$ become $n_t$, the paper's hours symbol, and the human capital shock becomes $\xi_{t+1}$, randomness in the law of motion. \citet{intertemporal_imai_2004} impose $c_t \ge 0$, $n_t \ge 0$ and $A_{t+1} \ge 0$. Because the horizon is finite and the primitives depend on age, the within-period theory applies at each age $t$ with that age's continuation value as $W$. To satisfy the main text's regularity assumption I bound the model as in the motivating model: $c_t \in [\underline{c}, \overline{c}]$ with $\underline{c} > 0$, $n_t \in [0, \overline{n}]$, and $A_t, A_{t+1} \in [0, \overline{A}]$. The shock $\xi_{t+1}$ has support $[\underline{\xi}, \overline{\xi}] \subset (0, \infty)$. Starting from a compact interval $[\underline{K}_0, \overline{K}_0]$ with $\underline{K}_0 > 0$, define $\underline{K}_{t+1} := \underline{\xi} \min K^{d}_{t+1}$ and $\overline{K}_{t+1} := \overline{\xi} \max K^{d}_{t+1}$, the extremes over $K_t \in [\underline{K}_t, \overline{K}_t]$ and $n_t \in [0, \overline{n}]$, and assume $\min K^{d}_{t+1} > 0$ at every age. The age-$t$ state space $\mathcal{X}_t = [0, \overline{A}] \times [\underline{K}_t, \overline{K}_t] \times \mathcal{S}_\varepsilon$, with $\mathcal{S}_\varepsilon$ the compact range of the disutility shock, is then compact and maps into $\mathcal{X}_{t+1}$ under the law of motion. Full-time earnings at the lowest human capital cover the consumption floor, $R_k \underline{K}_t\,\overline{n} \geq \underline{c}$ at every age. A consumption cap $\overline{c} \geq \overline{A} + R_k \overline{K}_t\,\overline{n}$ absorbs resources above the asset ceiling. Utility is bounded and continuous.

\textit{SDP Decomposition.} The within-period problem decomposes into two decision stages plus a terminal expectation step. I put labor supply and human capital investment first and consumption and saving second, so that hours fix the resources $m_t$ before the consumption stage allocates them. Following the recursive definition $v_j(\mathbf{y}_{j -1}) = \max_{\mathbf{a}_{j} \in \gamma_{j}} \{ u_{j}(\mathbf{y}_{j -1}, \mathbf{a}_{j}) + v_{j+1}(f_{j}(\mathbf{y}_{j -1}, \mathbf{a}_{j})) \}$, the stages are:

\textit{Stage 1 (Labor Supply and Human Capital).} The agent chooses labor supply $n_t$, determining period resources and future human capital.

\begin{itemize}
\item Input state: $\mathbf{y}_{0} = \mathbf{X}_t = (A_t, K_t, \varepsilon_t)$
\item Output state: $\mathbf{y}_{1} = (m_t, K^{d}_{t+1}, \varepsilon_t) \in \mathcal{Y}_1$, whose first two coordinates range over the compact image of $(A_t, K_t, n_t) \mapsto (m_t, K^{d}_{t+1})$ over $n_t \in [0, \overline{n}]$, with cash-on-hand in $[0, \overline{A} + R_k \overline{K}_t\,\overline{n}]$
\item Stage utility: $u_1 = -\varphi \varepsilon_{t} n_t^{\alpha_2}/\alpha_2$
\item Transitions: $m_t = A_t + R_k K_t n_t$ and $K^{d}_{t+1} = k_0 + \delta K_t + \omega(K_t, n_t, t)$
\item Stage set: $\gamma_1(A_t, K_t) = [\max\{0, (\underline{c} - A_t)/(R_k K_t)\}, \overline{n}]$, the hours that leave cash-on-hand at or above the consumption floor, nonempty and continuous because $R_k \underline{K}_t\,\overline{n} \geq \underline{c}$ and $K_t \geq \underline{K}_t > 0$

\begin{equation}
v_1(A_t, K_t, \varepsilon_{t}) = \max_{n_t \in \gamma_1(A_t, K_t)} \left\{ -\varphi \varepsilon_{t} \frac{n_t^{\alpha_2}}{\alpha_2} + v_2(m_t, K^{d}_{t+1}, \varepsilon_{t}) \right\}.
\end{equation}
\end{itemize}

\textit{Stage 2 (Consumption and Saving).} The agent chooses consumption $c_t$, determining end-of-period assets.

\begin{itemize}
\item Input state: $\mathbf{y}_{1} = (m_t, K^{d}_{t+1}, \varepsilon_t)$
\item Output state: $\mathbf{y}_{2} = (A_{t+1}, K^{d}_{t+1}, \varepsilon_t)$, the post-decision state
\item Stage utility: $u_2 = \Upsilon(t) \frac{c_t^{\alpha_1}}{\alpha_1}$
\item Transition: $A_{t+1} = (1+r)(m_t - c_t)$
\item Stage set: $\gamma_2(m_t) = [\max\{\underline{c}, m_t - \overline{A}/(1+r)\}, \min\{\overline{c}, m_t\}]$ at the reachable cash-on-hand $m_t \in [\underline{c}, \overline{A} + R_k \overline{K}_t\,\overline{n}]$, those that hours in $\gamma_1$ produce, nonempty because $m_t \geq \underline{c}$ and by the consumption cap

\begin{equation}
v_2(m_t, K^{d}_{t+1}, \varepsilon_{t}) = \max_{c_t \in \gamma_2(m_t)} \left\{ \Upsilon(t) \frac{c_t^{\alpha_1}}{\alpha_1} + v_3(A_{t+1}, K^{d}_{t+1}, \varepsilon_{t}) \right\}.
\end{equation}
\end{itemize}

\textit{Expectation step.} This step computes the expected continuation value from the post-decision state $\mathbf{y}_2$:
\begin{equation}
v_3(A_{t+1}, K^{d}_{t+1}, \varepsilon_{t}) = \beta \mathbb{E}_t [V_{t+1}(A_{t+1}, K^{d}_{t+1}\xi_{t+1}, \varepsilon_{t+1}) \mid \varepsilon_{t}].
\end{equation}

The disutility shock predicts $\varepsilon_{t+1}$, so it stays in every state; were it i.i.d., it would leave the state after stage 1, the only stage whose utility depends on it. With $J=2$ decision stages and $v_3 = W$, the decomposition satisfies (S1)-(S3).